\documentclass[onecolumn]{IEEEtran}
\usepackage{amsmath}
\usepackage{amssymb}
\usepackage{graphicx} 
\newtheorem{theorem}{Theorem}
\newtheorem{lemma}{Lemma}
\newtheorem{remark}{Remark}
\newtheorem{definition}{Definition}

\usepackage{color}

\IEEEoverridecommandlockouts 
\begin{document}
\title{Representation and Network Synthesis  for a Class of Mixed Quantum-Classical Linear  Stochastic  Systems}
\author{Shi Wang~~ Hendra I. Nurdin~~Guofeng Zhang~~Matthew R. James\thanks{S.~Wang is with College of Electrical and Information
Engineering, Hunan University, Changsha, 410082, China. Hendra I. Nurdin is with School of
    Electrical Engineering and Telecommunications, University of New South Wales, Sydney,
N.S.W. 2052
Australia. Guofeng Zhang is with Department of Applied Mathematics,
The Hong Kong Polytechnic University, Hung Hom, Kowloon,
HKSAR, China. Matthew R. James is with Centre for Quantum
Computation and Communication Technology, Research  School of
    Engineering, Australian
    National University, Canberra, ACT 0200,
    Australia.}}
\maketitle
\begin{abstract}
The purpose of this paper is to present a network realization theory for a class of mixed quantum-classical linear stochastic systems. Two forms, the standard form and the general form, of this class of linear mixed quantum-classical systems are proposed. Necessary and sufficient conditions for their physical realizability are derived. Based on these physical realizability conditions, a network synthesis theory for this class of linear mixed quantum-classical systems is developed, which clearly exhibits the quantum component, the classical component, and their interface. An example is used to illustrate the theory presented in this paper.
\end{abstract}

\section{Introduction}
\label{sec:intro}
Linear systems  are of
basic importance to classical control engineering, and also arise in the
modeling and control of quantum systems; see, e.g., \cite{GZ04}, \cite{DA2007}, \cite{MvH07}, \cite{GJ09a}, \cite{SHHPJ09},  \cite{WM10}, \cite{DP10}, \cite{ZJ11}, \cite{ZJ12}, \cite{AT12}, \cite{JK2014}, \cite{GZ15}, \cite{WSPSGK15},  \cite{ZLW+17}, \cite{XPD17}, \cite{NY17}, \cite{ZGPG18}. A classical  linear system described by the state space
representation can be realized using electrical
components by linear electrical network synthesis theory, see
\cite{AV1973}. Linear quantum optical systems may be described by
linear quantum differential equations in the Heisenberg picture of
quantum mechanics, \cite{GZ04}, \cite{JNP08}, \cite{NJP09}, \cite{WM10},  \cite{GJ09a}, \cite{WNZJ13}, \cite{GZ15}, \cite{WSPSGK15}, \cite{NY17}, \cite{ZGPG18}.   Such quantum  linear systems described by the state space
representation can be built by optical cavities, degenerate parametric
amplifiers (DPA), phase shifters, beam splitters, and squeezers,
etc;  interested readers may refer to \cite{Leo03}, \cite{BR04}, \cite{NJD09}, \cite{NY17} for
a more detailed introduction to these optical devices. Quantum technologies often comprise quantum systems interconnected with classical (non-quantum) devices, which means that the two types of  systems may be connected as an integral whole (called mixed quantum-classical systems in this paper) by appropriate interfaces that  convert quantum signals to classical signals, and vice-versa. Traditionally,
such quantum optical networks would be implemented on an optical
table. However, it is now becoming possible to consider
implementation in semiconductor chips, \cite{BKSWW2007}, \cite{OBFV09}, \cite{WNZJ13}.

\begin{figure}[htbp]
\centering
\includegraphics[scale=0.39]{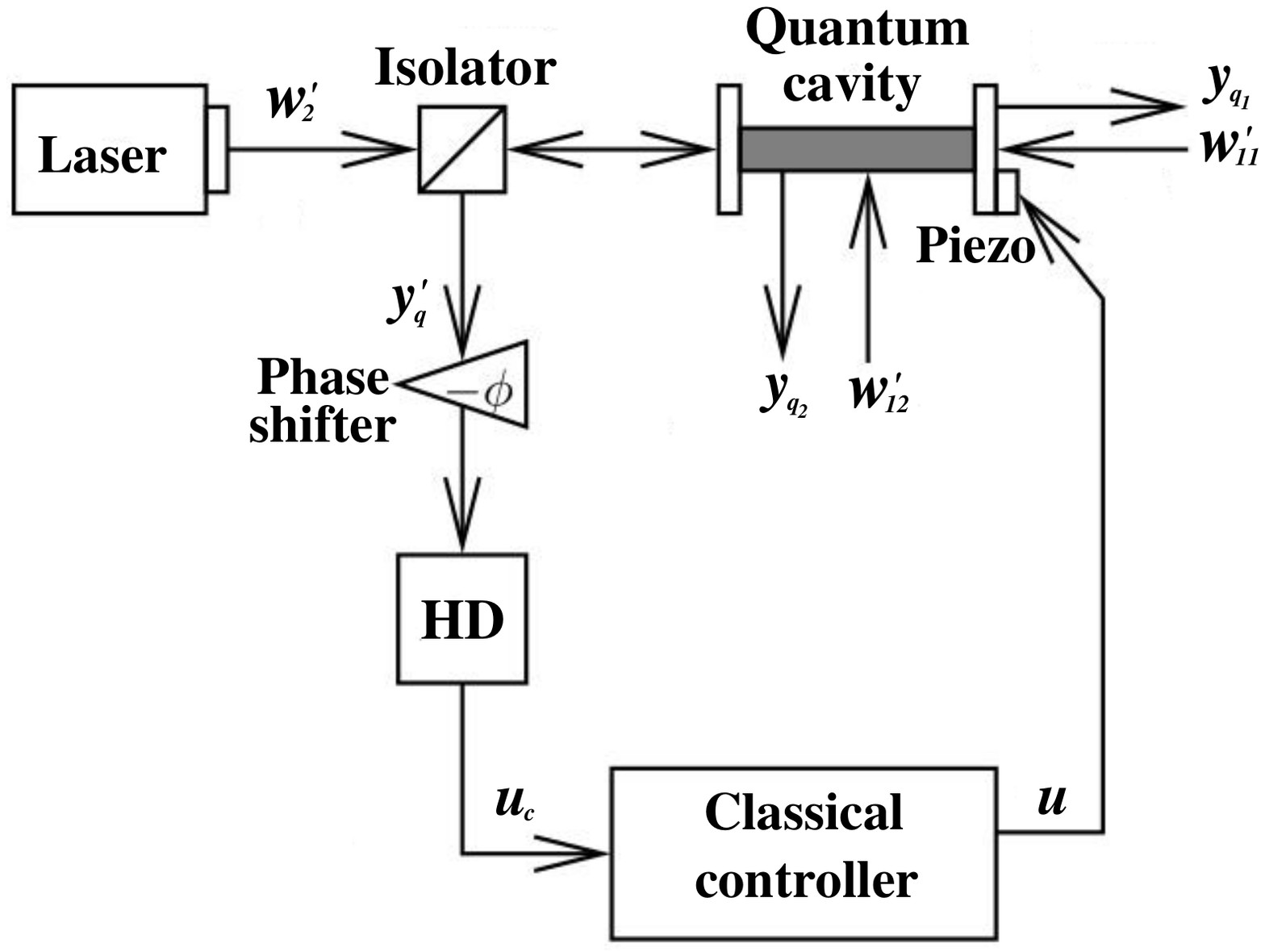}
\vspace{0em}\caption{A mixed quantum-classical system (Cavity locking feedback control loop) studied in \cite{SHHPJ09}. }
\label{fig:realnew}
\end{figure}

In classical control engineering,  many methods have been developed for designing controllers that meet various control specifications. The design process
begins with some form of specification for the system, and concludes with a physical realization of the controller that meets the
specifications. Often, mathematical models for the controller are used
in the design process, such as state space equations for the controller.
These state space equations may result from a mathematical
optimization procedure, such as $H^\infty$, LQG, or some other procedure. The
process of going from such mathematical models to the desired
physical systems is a process of {\em synthesis} or {\em physical
realization}, part of the design methodologies widely used in
classical engineering \cite{AV1973}. Analogous design issues are beginning to present themselves in
quantum technology. A quantum control system often has both quantum  and classical components. Indeed, in measurement-based feedback control, a classical controller is used to control a quantum plant. That is, a quantum control system is often a mixed quantum-classical system.  Figure
\ref{fig:realnew} illustrates an example of a mixed
quantum-classical linear system studied in \cite{SHHPJ09}. In this measurement-based feedback control system,  a Fabry-Perot optical cavity \cite{BR04}, \cite{NJD09}, \cite{WM08}, which is described quantum-mechanically,  is
connected to a classical controller via a homodyne detector (HD) and a  piezo-electric actuator \cite{WM10}, \cite{WM93}.  The light field (quantum signal) reflected
from the cavity is first separated from the incoming laser beam by an optical
isolator, and then is detected by a HD (a quantum-to-classical converter), thus yielding a photocurrent which is a classical signal. The classical
controller processes such classical signals to generate a classical control input $u$, which is then fed back to regulate the  optical path length of the cavity via the piezo-electric actuator  in order to actuate the resonant frequency of the cavity.   Interested reader may refer to \cite{SHHPJ09} for more details.  

The purpose of this paper is to propose canonical representations for  a class of linear stochastic
differential equations that may describe mixed quantum-classical
systems and then develop a network synthesis theory for such class of equations that
reveals in a  clear way the internal structure of a
mixed quantum-classical system. Furthermore, arbitrary linear
stochastic differential equations for mixed systems need not correspond to a physical system, and so we derive
conditions ensuring that they do; that is, physical realizability. This work generalizes and extends earlier work \cite{JNP08},
\cite{Nurdin2011}, \cite{WNZJ2012}. In \cite{WNZJ2012}, we only consider a {\em standard} model for mixed quantum-classical linear stochastic systems for the design process. However, in this paper, we will investigate a more general model for the physical realization of  mixed quantum-classical linear stochastic systems.

The rest of the paper is organized as follows. Section \ref{sec:pre} introduces
some concepts about classical and quantum random variables as well as probabilities,  briefly describes closed  quantum harmonic oscillators, and also gives a brief overview of  linear non-commutative stochastic systems and
non-demolition conditions. Section \ref{sec:mod} proposes two models
of mixed quantum-classical linear stochastic systems for the design
process and presents a connection between these models. Section
\ref{sec:realization} presents physical realizability definitions
and constraints for the two models defined in Section \ref{sec:mod},
respectively. Section \ref{sec:them} develops a network synthesis
theory for a mixed quantum-classical system. Section \ref{sec:application}
presents a potential application of the main results of Section \ref{sec:them}.  Finally, Section \ref{sec:conclusion} gives the
conclusion of this paper.

\section{Preliminaries}\label{sec:pre}

\subsection{Notation}
The notations used in this paper are as follows. The imaginary unit is
$i=\sqrt{-1}$. The commutator of two operators $A$ and $B$
is defined by $[A, ~B]=AB-BA$. If $x$ and $y$ are column vectors of
operators, the commutator is defined by $[x, ~y^T]= xy^T -(yx^T)^T$.
If $X=[x_{jk}]$ is a matrix of linear operators or complex numbers,
then
$X^{\#} = [x^{*}_{jk}]$ denotes
the operation of taking the adjoint of each element of $X$, and
$X^{\dag}=[x^{*}_{jk}]^T$. We
also define $\Re(X)=(X + X^{\#})/2$ and $\Im(X)=(X -X^{\#})/2i$.  The symbol $I_k$ denotes the $k \times k$
identity matrix, $0_{j\times k}$ denotes the $j\times k$ zero
matrix and $0_k \equiv 0_{k\times k}$.  Let
$J=\left[
\begin{array}{cc}
0 & 1 \\
-1 & 0 \\
\end{array}
\right]$ and
$\mathrm{diag}_{k}(M)$ denote
a block diagonal matrix with the square matrix $M$ appearing $k$ times
on the diagonal block. A symplectic matrix $V$ of dimension $2k \times 2k$ is a real matrix satisfying $V\Theta_k V^T = \Theta_k$, where $ \Theta_k = {\rm diag}_k (J)$. We set $\hbar =1$ throughout this paper.

\subsection{Classical and quantum random variables}
\label{sec:preliminaries-random}

A classical random variable, usually written as $X$, is a variable whose possible values are numerical outcomes of a random phenomenon. A random variable $X$ with mean  $\mu = \mathbf{E}[ X]$ and variance $\sigma^2 =
\mathbf{E}[ (X-\mu)^2 ]$ is said to be {\em
Gaussian} if its probability distribution function $F$ is Gaussian,
i.e.,
\begin{eqnarray} \label{jun29_F}
F (a <  X \leq b ) = \int_a^b p_X(x) dx,  ~~~ \forall -\infty < a<b < \infty,
\end{eqnarray}
where $p_X(x) =
\frac{1}{\sigma\sqrt{2\pi}}\mathrm{exp}(-\frac{(x-\mu)^{2}}{2\sigma^2})$ is of course the well-known Gaussian probability density function.

In quantum physics, a quantum random variable $A$ is an operator defined on a Hilbert space $\mathfrak{H}$. In particular, if $A$ is self-adjoint, it is called  an  {\it observable}   and can be used to represent
some physical quantity. Because an observable is self-adjoint, by the spectral theory, its spectra are real numbers.  Actually, an observable can be physically measured to generate outcomes which are real numbers. On the other hand,  a quantum {\it state}   $\psi$  encodes an experimenter's knowledge or information about some aspect of reality and  is given mathematically as a vector of $\mathfrak{H}$, permitting the calculation of  expected values of quantum random variables.  If an observable $A$  is measured on a quantum system  prepared in the state  $\psi$, then its mean value is given by the inner product  $\langle \psi, A
\psi \rangle=\int_{-\infty}^{\infty} \psi(q)^* A \psi(q)dq$. In quantum mechanics,  the Dirac ``ket'' notation $|\psi\rangle$ is always used to denote a  pure quantum state $\psi$. The adjoint of $|\psi\rangle$ is the ``bra'' vector $\langle \psi|$. Then,  we can write the previous inner product as $\langle \psi, A
\psi \rangle = \langle \psi | A
\psi \rangle$. Moreover, we can associate a density operator $\rho$ with state $|\psi\rangle$ as $\rho = |\psi\rangle \langle \psi|$. The density operator $\rho$ defined in this way corresponds to a pure state and is a rank-1 projector, but in general $\rho$ can also be used to describe a classical mixture of pure states, \cite{EM98,WM10}.

Consider an example of a quantum harmonic oscillator with  amplitude quadrature operator $Q$ and phase quadrature operator $P$, a model for an optical mode in a cavity. The two observables $Q$ and $P$ are defined by
\begin{eqnarray} \label{eq:QP_1d}
(Q \psi)(q) = q \psi (q), \ \ \ (P\psi)(q) = -i    \frac{d}{dq}
\psi(q)
\end{eqnarray}
for $\psi \in \mathfrak{H} = L^2(\mathbb{R})$, respectively.   In quantum mechanics, the amplitude and phase quadrature
 observables satisfy the commutation
relation $[Q,~P]=2i$, and such non-commuting
observables are referred to as being {\it incompatible}. The state vector
\begin{eqnarray}
\psi(q) =
(2\pi)^{-\frac{1}{4}}\sigma^{-\frac{1}{2}}\mathrm{exp}\left(-\frac{(q-\mu)^{2}}{4\sigma^2}\right)
\end{eqnarray}
is an instance of what is known as a quantum {\em Gaussian state}. For this
particular {\em Gaussian state}, the means of $P$ and $Q$ are given by
$\int_{-\infty}^{\infty} \psi(q)^* Q\psi(q)dq =\mu$, and
$\int_{-\infty}^{\infty} \psi(q)^* P\psi(q)dq =0$, and similarly the
variances are $\sigma^2$ and $\frac{1}{4\sigma^2}$,
respectively.

\subsection{Classical probability and quantum probability}\label{sec:preliminaries-probability}

In the classical probability theory, a  probability model is given by a triple ($\Omega, F, \nu$), where
 \begin{enumerate}
\item the sample space $\Omega$ is the set of all possible outcomes of some experiment;

\item $F$ is a collection of events, which are subsets of $\Omega$;

\item $\nu$ is a probability measure.
\end{enumerate}
Classical random variables can be defined on the probability space ($\Omega, F, \nu$). For instance, when $\Omega = \Bbb{R}$, $F$ is the $\sigma$-field generated by all the sets of the form $(a,b]$, and probability measure $\nu$ is defined in terms of  $p_X(x)$ used given in Eq. \eqref{jun29_F}, specifically,
\[
\nu(A) = \int_{A} p_X(x)dx,~~~ \forall A \in F,
\]
then the associated random variable $X$ is a Gaussian random variable.

The  quantum probability model,  a generalization of   the classical
probability model \cite{AFL82}, \cite{HP84}, \cite{BHJ07}, can be defined  at the level of von Neumann algebra and density operators. More specifically, a quantum probability model ($\mathcal{A}, \rho$) (also called a quantum probability space) consists of
\begin{enumerate}
 \item a von Neumann algebra $\mathcal{A}$ generated by a collection of  projection operators on a Hilbert space $\mathfrak{H}$ (the projections $E\in \mathcal{A}$  are called events in $\mathcal{A}$);
 \item a density operator $\rho$. The trace ${\rm tr}[\rho E]$ gives the probability that an event $E\in \mathcal{A}$ occurs, where  events in a quaantum probability space are represented by projection operators.
\end{enumerate}


The quantum probability model is the most
natural non-commutative generalization of classical
probability, in the sense that every classical probability space can be embedded in a quantum probability space. For example,  given a vector of classical  Gaussian random  variables
$\tilde{X}=[X_1\quad X_2\quad\cdots\quad X_k]^T$ with joint probability density
function
\begin{eqnarray}
f(q) =
(2\pi)^{-\frac{k}{2}}|\Sigma|^{-\frac{1}{2}}\mathrm{exp}\left(-\frac{1}{2}(q-\tilde{\mu})^T\Sigma^{-1}(q-\tilde{\mu})\right)
\end{eqnarray}
with mean $\tilde{\mu}\in\mathbb{R}^{k}$ and covariance matrix
$\Sigma\in\mathbb{R}^{k\times k}$, we may define the quantum state $\psi= \sqrt{f(q)}$ and   a vector of  quantum observables $\check X_Q=[Q_1\quad Q_2\quad\cdots\quad Q_k]^T$   by means of $(Q_j\psi)(q) = q_j \psi(q)$ for $j=1,\ldots,k$. It is easy to show that the classical
Gaussian random  variables $\tilde{X}$ and the quantum Gaussian random variables $\check X_Q$ have the same mean and variance values, thus having  the identical distribution. So statistically,
$\tilde{X} \equiv \check X_Q$. In the similar way as in Eq. (\ref{eq:QP_1d}),  we may define a vector of quantum observables $\check X_P=[P_1\quad
P_2\quad\cdots\quad P_k]^T$ by means of $(P_j\psi)(q) = -i    \frac{\partial}{\partial q_j}\psi(q)
$ for $j=1,\ldots,k$. Then it is easy to see that $[\check X_Q, ~ \check X_P^T]=2i  I_k$.      Let $\mathbf{P}\in\mathbb{R}^{2k\times 2k}$ be a permutation matrix such that $\mathbf{P} \check
X = [Q_1 \quad P_1 \quad\cdots\quad Q_k \quad P_k]^T$. Then the commutation relation becomes $[(\mathbf{P}\check
X), ~(\mathbf{P}\check X)^T] =2i{\rm diag}_k (J)\equiv 2i\Theta_k$. The quantum vector $\check
X=[\check X_Q^T\quad \check X_P^T]^T$ is called an {\em
augmentation} of $\tilde{X}$. The relation between classical and quantum random variables may be
expressed as
$
\tilde{X }\equiv [ \begin{array}{cc} I_k & 0_k
\end{array} ]
\left[  \begin{array}{c} \check X_Q\\ \check X_P
\end{array} \right]$. In the rest of the paper, we use symbol ``$=$'' instead of ``$\equiv$'' to represent such equivalence relation. However, ``$=$'' here only means that the classical random variable  $\tilde{X }$ and the quantum observable $\check X_Q$ have the same probability distribution. Recall that the probability distribution $\nu_Q$ of $\check X_Q$ can be defined as follows. Let $E_{Q_j}$ be the spectral measure of $Q_j$ (i.e., the projection operator-valued measure such that $Q_j (A)= \int_{A} x E_{Q_j}(dx)$ for all Borel subsets $A$ of $\mathbb{R}$). Let $ B(\mathbb{R}^j)$ denote the $\sigma$-algebra generated by the Borel subsets of $\mathbb{R}^j$ for any positive integer $j$. Then the probability distribution of $\check X_Q$ is the probability measure $\nu_Q$ on the measurable space $(\mathbb{R}^n,B(\mathbb{R}^n))$  defined as $\nu_Q(A_1 \times A_2 \times \cdots \times A_j) = {\rm tr}(\rho \prod_{j=1}^k E_{Q_j}(A_j))$ for any $A_l \in B(\mathbb{R})$, $l=1,\ldots,k$ and then uniquely extended to a probability measure on $(\mathbb{R}^k,B(\mathbb{R}^k)$). Note that $E_{Q_j}(A)$ and $E_{Q_k}(B)$ commute for any $j \neq k$ and any $A,B$, and $E_{Q_j}(A_j)$ under the trace should be interpreted as the amplitation of $E_Q(A_j)$ to a projection operator on the composite Hilbert space of the $k$ oscillators.

\subsection{Classical linear stochastic systems}
As is well known, in control engineering, a state-space representation is a mathematical model of a physical system as a set of input, output and state variables  described by a set of  ordinary differential equation. Consider  a classical linear  system $G_1$  given in a state space representation which may describe an electrical or electronic circuit as:
\begin{eqnarray}
dx_c(t)&=&A_{cc}'x_c(t)dt+B_c'du_c(t)\label{classical-eq1},\\
dy_c(t)&=&C_{cc}'x_c(t)dt+D_c'du_c(t)\label{classical-eq2},\\
y_{c1}'(t)&=&C'_{c_1}x_c(t), \quad
y_{c2}'(t)=C'_{c_2}x_c(t),~~~ t\geq 0.
\label{classical-eq4}
\end{eqnarray}
Here, $x_c(t)$ represents a vector of $n_c$ classical variables; $y_{c}$, $y_{c1}'$ and $y_{c2}'$ are vectors of classical output signals of dimension $n_{y_c}$, $2n_{w_1}$, and $2n_{w_2}$ respectively\footnote{As the classical system $G_1$ will become part of the mixed quantum-classical system \eqref{standard-form_a}-\eqref{standard-form_b} , we specify the numbers of system variables and outputs  for future use. The number of inputs will be given later. Moreover, the superscript $'$ indicates that these matrices or outputs are for interconnections. Similar convention is used for the quantum system $G_2$ to be given in Eqs.  \eqref{quantum-eq1}-\eqref{quantum-eq3}.}. The classical input signal $u_c(t)$ has the form $du_c(t)=\alpha_c(t)dt + dw_c(t)$, where $w_c(t)$ is a vector of independent standard classical Wiener processes, and $\alpha_c(t)$ is a vector of real stochastic processes of locally bounded variation. $A_{cc}'$, $B_c'$, $C_{cc}'$, $D_c'$, $C'_{c_1}$ and $C'_{c_2}$ are all real constant matrices.

\subsection{Quantum  linear stochastic systems and physical examples}\label{subsec:C-Q-systems}
In this subsection, we will introduce some basic examples of physical systems that are linear quantum stochastic systems, coming from the field of quantum optics. At the end of the section we then provide a description of a general class of linear quantum stochastic systems.

Before  presenting the basic examples, we start with a model of a {\em closed quantum harmonic oscillator} which may help readers better understand our  models proposed later in the paper. For a more detailed exposition, we refer to \cite{Gardiner85,GZ04,GJ09a,NY17}.

\subsubsection{Closed quantum harmonic oscillator}
A quantum harmonic oscillator is said to be a {\em closed quantum harmonic oscillator} if it is completely isolated  from any external environment. In other words, it does not interact with an environment and evolves only under its own Hamiltonian. Now we describe the dynamics of a closed quantum harmonic oscillator with   position and momentum operators $Q$ and $P$ as  defined in Subsection \ref{sec:preliminaries-random}. Its Hamiltonian $H_o$ is given by
\begin{equation}\label{Hamiltonian1}
 H_o={\frac {{P}^{2}}{2m}}+{\frac {1}{2}}m\omega ^{2}{Q}^{2},
\end{equation}
where $m$ is the oscillator's mass and $\omega$ is the angular frequency of the oscillator. From \eqref{Hamiltonian1}, we have the Heisenberg equations of motion for  $Q$ and $P$ given by
\begin{equation}
\left[
  \begin{array}{c}
    \frac{dQ}{dt}  \\
   \frac{dP}{dt} \\
  \end{array}
\right]
=\left[
               \begin{array}{cc}
                0 & \frac{1}{m} \\
                 -m\omega^2 & 0 \\
               \end{array}
             \right]\left[
  \begin{array}{c}
   Q\\
 P \\
  \end{array}
\right].
\end{equation}
Therefore, $Q(t)= \mathrm{cos}(\omega t)Q(0)-\mathrm{sin}(\omega t)P(0)$,\ \ $P(t) =\mathrm{sin}(\omega t)Q(0)+\mathrm{cos}(\omega t)P(0)$. Next we we will allow quantum harmonic oscillators to interact with  electromagnetic fields to produce
open quantum systems. The dynamical behavior of open quantum systems plays a key role in many applications of quantum mechanics.

\subsubsection{Quantum fields and examples of open quantum optical systems}
No quantum system is completely isolated from its environment.  The quantum system is said to be an {\em open quantum system} if it is interacting with an environment. In particular, in quantum optics this environment can take the form of an external electromagnetic (EM) field, which is a boson field.

Under some physical assumptions regarding the interaction of the  field and the oscillator, the field can be modelled as an operator-valued white noise field $b(t)$ satisfying the singular commutation relation $[b(t),b^*(s)]=\delta(t-s)$. These assumptions can include a combination of rotating wave approximation and the Markov assumption, or the weak coupling limit between the oscillator and the field and coarse graining of time, depending on the system being considered. For a detailed discussion of these assumptions and their physical motivations, we refer the reader to the seminal contributions of Gardiner and Collett \cite{Gardiner85} and  the physics text \cite[Chapters 3, 5 and 10]{GZ04} and \cite{ALV02}. The class of models described herein are widely accepted as highly accurate models for linear quantum optical devices as well as for devices from other related domains such as optomechanics and microwave superconducting circuits. The interaction Hamiltonian $H_{\rm int}(t)$ between the oscillator and the field, given in the interaction picture with respect to the free-field dynamics, takes the form
$$
H_{\rm int}(t) = -i (b(t)^*L  - L^* b(t)),
$$
where $L$ is an operator of the oscillator. Let
$$
a = \frac{Q+iP}{2}$$
be the annihilation operator for the oscillator, satisfying the commutation relation $[a,a^*]=1$. For the concrete examples below $L$ takes on the form $L= \sqrt{\kappa} a$, where $\kappa$ is a constant called the decay rate and $a$ is the annihilation operator of the field.

\begin{figure}[htbp]
\vspace{-2em}\centering
\includegraphics[height=50mm,width=70mm]{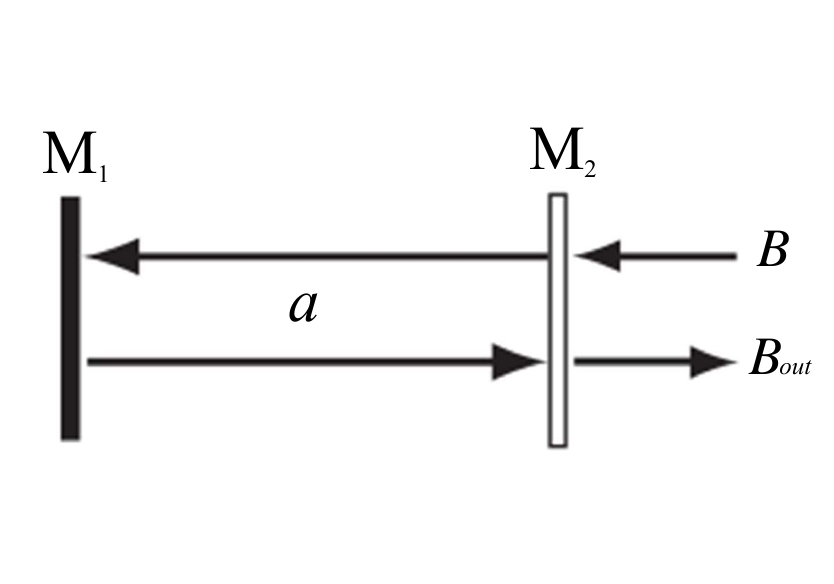}
\vspace{-2em}\caption{Open optical cavity. The cavity consists of  two mirrors: $\mathrm{M_1}$ denotes a fully reflecting
mirror and $\mathrm{M_2}$ denotes a partially transmitting mirror. \label{fig:l1}}
\end{figure}

1) {\em Optical Cavity}:
Consider a single open optical cavity as shown in Figure \ref{fig:l1}. This type of cavity is known as a Fabry-Perot cavity with a mode corresponding to a standing light field formed between the mirrors $\mathrm{M}_1$ and $\mathrm{M}_2$ by the bouncing of light back and forth between them.
The cavity alone is modeled by a single quantum harmonic oscillator with Hamiltonian $H_{\rm cav}=\omega_{cav}a^{*}a$ with the  resonance frequency $\omega_{cav}$ and  the cavity annihilation operator $a$ is as defined before. However, the optical cavity in the figure will interact with an external EM field through the mirror $\mathrm{M}_2$, therefore it is an open quantum system. At this mirror there can be an exchange of photon between the cavity and the external field. It is convenient to work with the integrated version of the white noise field $B(t)= \int_{0}^t b(s) ds= (w_q(t)+i w_p(t))/2$, where $w_q(t)$ and $w_p(t)$ are self-adjoint non-commuting quantum Wiener processes.   These processes can be realized as on symmetric Fock space over the space of square integrable complex functions \cite{HP84,GJ09a}. We remark that each of the processes $w_q(t)$ and $w_p(t)$ are individually isomorphic to a classical Wiener process but, since they do not commute, they cannot be realized on a common classical probability space.

As alluded to earlier, for the optical cavity, $L= \sqrt{\kappa} a$, where $a$ is the cavity annihilation operator. The joint dynamics of an open optical cavity coupled to the bosonic field $B(t)$ may be described by a unitary propagator $U(t)$ satisfying the quantum stochastic differential equation \cite{HP84,WM08},
$$
dU(t) = \left( -(i H_{\rm cav} +\frac{\kappa}{2} a^*a)dt + \sqrt{\gamma} dB^*(t)a - \sqrt{\gamma} a^* dB(t) \right) U(t),
$$
with initial condition $U(0)=1$. For each $t \geq 0$, the solution of this equation is unitary, $U(t)^*U(t)=U(t)U(t)^*=I$. The Heisenberg picture evolution of the cavity's annihilation operator $a$ and creation operator $a^*$, are respectively, $a(t) = U(t)^* a U(t)$ and $a^*(t) = U(t)^* a U(t)$. They satisfy the linear quantum stochastic differential equation, \cite{GZ04,GJ09a}:
\begin{eqnarray}
\left[
  \begin{array}{c}
    da(t)\\
    da^*(t)\\
  \end{array}
\right]&=&\left[
            \begin{array}{cc}
              -\frac{\kappa}{2}-i\omega_{cav} & 0 \\
              0 & -\frac{\kappa}{2}+i\omega_{cav} \\
            \end{array}
          \right]
\left[
  \begin{array}{c}
    a(t)\\
    a^*(t)\\
  \end{array}
\right]dt-\sqrt{\kappa}\left[
  \begin{array}{c}
   dB(t)\\
    dB^*(t)\\
  \end{array}
\right] \label{cavity1}.
\end{eqnarray}
After interacting with the cavity the field $B(t)$ also undergoes a transformation in the Heisenberg picture, yielding the output field $B_{\rm out}(t) = U(t)^* B(t) U(t)$. This is the field reflected from the cavity that contains any photons that have escaped from the cavity through the mirror $\mathrm{M}_2$, see Fig.~\ref{fig:l1}. The output field $B_{\rm out}(t)$ satisfies the output equation:
\begin{eqnarray}
dB_{out}(t)&=&\sqrt{\kappa}a(t)dt+dB(t). \label{cavity1-out}
\end{eqnarray}

\begin{figure}[htbp]
\vspace{-1em}\centering
\includegraphics[height=50mm,width=95mm]{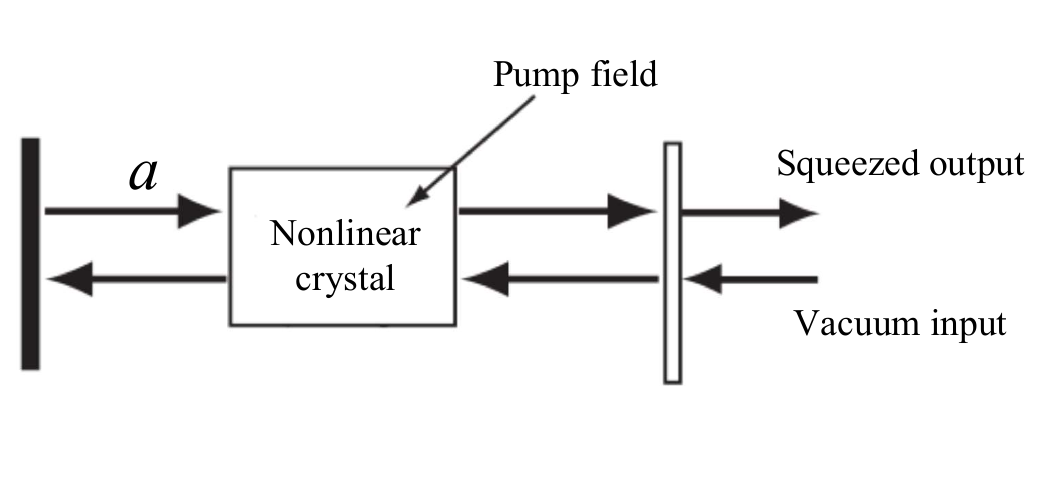}
\vspace{-2em}\caption{Degenerate parametric amplifier coupled to an bosonic field. \label{fig:l2}}
\end{figure}

2) {\em Degenerate Parametric Amplifier}: Now we briefly describe a degenerate parametric amplifier (DPA) as shown in Figure \ref{fig:l2}. It is an open oscillator with a classical pump that can produce squeezed output field (a field with reduced fluctuations along one of its quadratures and increased fluctuations on the conjugate quadrature). The pump field provides quanta and interacts with the cavity mode in a type of crystal called a $\chi^{(2)}$ crystal. In this crystal one photon from the pump field is annihilated to produce two photons of the cavity mode. The amplifier's Hamiltonian can be written as $$H_{SQ}=\omega_{cav} a^{*}a+\frac{i}{2}(\epsilon e^{-i\omega_p t} a^{*2}-\epsilon^* e^{i\omega_p t}a^2),$$ where $\omega_p$ is the frequency of the pump beam and $\epsilon$ a measure of the effective pump amplitude.
Following the same procedure as for the optical cavity, the Heisenberg picture dynamics of a degenerate parametric amplifier coupled to a  bosonic field $B(t)$ is now given by (see \cite[Chapter 10]{GZ04})
\begin{eqnarray*}
\left[
  \begin{array}{c}
    da(t)\\
    da^*(t)\\
  \end{array}
\right]&=&\left[
            \begin{array}{cc}
              -\frac{\kappa}{2}-i\omega_{cav} & \epsilon e^{-i\omega_p t} \\
              \epsilon^* e^{i\omega_p t} & -\frac{\kappa}{2}+i\omega_{cav} \\
            \end{array}
          \right]
\left[
  \begin{array}{c}
    a(t)\\
    a^*(t)\\
  \end{array}
\right]dt-\sqrt{\kappa}\left[
  \begin{array}{c}
   dB(t)\\
    dB^*(t)\\
  \end{array}
\right],\nonumber\\
dB_{out}(t)&=&\sqrt{\kappa}a(t)dt+dB(t).
\end{eqnarray*}
If we now switch to a rotating frame at half of the pump beam frequency $\omega_p/2$, we can remove the time-dependence in the system matrices to transform the system into a time-invariant one.  This entails making  the substitution $a(t) \rightarrow a(t)e^{-i\omega_p t/2}$, $a^*(t) \rightarrow a^*(t)e^{i\omega_p t/2}$, $B(t) \rightarrow B(t)e^{-i \omega_p t/2}$, $B^*(t) \rightarrow B^*(t)e^{i \omega_p t/2}$, and $B_{\rm out}(t) \rightarrow B_{\rm out}(t)e^{-i \omega_p t/2}$. These substitutions yield the time-invariant equation,
\begin{eqnarray}
\left[
  \begin{array}{c}
    da(t)\\
    da^*(t)\\
  \end{array}
\right]&=&\left[
            \begin{array}{cc}
              -\frac{\kappa}{2}-i(\omega_{cav} -\omega_p/2) & \epsilon  \\
              \epsilon^* & -\frac{\kappa}{2}+i(\omega_{cav}-\omega_p/2) \\
            \end{array}
          \right]
\left[
  \begin{array}{c}
    a(t)\\
    a^*(t)\\
  \end{array}
\right]dt-\sqrt{\kappa}\left[
  \begin{array}{c}
   dB(t)\\
    dB^*(t)\\
  \end{array}
\right],\nonumber\\
dB_{out}(t)&=&\sqrt{\kappa}a(t)dt+dB(t). \label{degenerate1}
\end{eqnarray}
The output field of a degenerate parametric amplifier as shown in Fig.~\ref{fig:l2} will be a squeezed field.

\subsubsection{More general model of open quantum stochastic systems}\label{subsection2.5}

It can be seen from from \eqref{cavity1}-\eqref{cavity1-out} and \eqref{degenerate1},  that  the system coefficients are complex. However, for solving the control  engineering problems, it is sometimes more convenient to work with systems with real-valued coefficients. Using the relations $x_{q_1}=Q(t)=a(t)+a^*(t)$, $x_{q_2}=P(t)=-i\left(a(t)-a^*(t)\right)$, $w_1'(t)=B(t)+B^*(t)$, $w_2'(t)=-i\left(B(t)-B^*(t)\right)$, $y_{q_1}=B_{out}(t)+B_{out}^*(t)$ and $y_{q_2}=-i\left(B_{out}(t)-B_{out}^*(t)\right)$, we can rewrite \eqref{cavity1} and  \eqref{cavity1-out}  in the quadrature representation (all system coefficients are real) as follows
\begin{eqnarray}
dx_q(t)&=&\left[
            \begin{array}{cc}
              -\frac{\kappa}{2} & \omega_{cav} \\
              -\omega_{cav} & -\frac{\kappa}{2} \\
            \end{array}
          \right]x_q(t)dt-\sqrt{\kappa}dw'(t)\nonumber,\\
dy_q(t)&=&\sqrt{\kappa}x_q(t)dt+dw'(t)\label{cavity-eq2},~~~ t\geq 0, \label{cavity-eq2}
\end{eqnarray}where $x_q(t)=\left[
                              \begin{array}{c}
                                x_{q_1} \\
                                x_{q_2} \\
                              \end{array}
                            \right]
$, $y_q(t)=\left[
                              \begin{array}{c}
                                y_{q_1} \\
                                y_{q_2} \\
                              \end{array}
                            \right]
$ and $w'(t)=\left[
                              \begin{array}{c}
                                w_1' \\
                                 w_2' \\
                              \end{array}
                            \right]$.

Similarly,  \eqref{degenerate1} can be rewritten as
\begin{eqnarray}
dx_q(t)&=&\left[
            \begin{array}{cc}
              -\frac{\kappa}{2}+ \Re\{\epsilon\} & \quad\omega_{cav} - \omega_p/2 + \Im\{\epsilon\}  \\
              -\omega_{cav} +\omega_p/2 + \Im\{\epsilon\} & \quad-\frac{\kappa}{2}-\Re\{\epsilon\}  \\
            \end{array}
          \right]x_q(t)dt-\sqrt{\kappa}dw'(t)\nonumber,\\
dy_q(t)&=&\sqrt{\kappa}x_q(t)dt+dw'(t)\label{cavity-eq2},~~~ t\geq 0. \label{cavity-eq2}
\end{eqnarray}

The optical cavity and degenerate parametric amplifier that we have briefly illustrated above are two important examples of linear quantum stochastic systems. We are now in the position to present a more general model for an open linear stochastic quantum system denoted by $G_2$ in the quadrature form  given  by
\begin{eqnarray}
dx_q(t)&=&A_{qq}x_q(t)dt+B_{q}dw'(t)+Eu(t)dt\label{quantum-eq1},\\
dy_q(t)&=&C_{qq}x_q(t)dt+D_qdw'(t)\label{quantum-eq2},\\
dy_q'(t)&=&C'_{qq}x_q(t)dt+D'_qdw'(t),~~~ t\geq 0, \label{quantum-eq3}
\end{eqnarray}
where $x_q(t)$ denotes $n_q$ pairs of  amplitude and phase quadrature  operators defined on a Hilbert space $\mathcal{H}$,
$w'(t)$ is a vector of $2m$ quantum stochastic  processes that can be represented as self-adjont operators defined on  a Fock space $F$, while $y_q(t)$ is a vector of $2n_{y_q}$ quantum outputs and $y_q'(t)$ is  a vector of $2n_{y_q'}$ quantum outputs, such that $n_{y_q} + n_{y_q'}\leq m$.

By the laws of quantum mechanics,  the quantum system $G_2$ is required to possess the following properties; see \cite[Section 2.5]{NY17}, \cite{GJN2010} for a more detailed discussion.

\begin{description}
\item[(a)]  The system variables preserve commutation relations, \cite{JNP08}:
\begin{equation} \label{xq_jun14}
 [ x_{q}(t), ~x_{q}(t)^T] =  [ x_q(0), ~x_q(0)^T]= 2 i \Theta_{n_q} , ~~ t\geq 0,
\end{equation}
where  $\Theta_{n_q}$ is a skew-symmetric real matrix. Moreover,  the
matrix $\Theta_{n_q}$ is said to be {\em canonical} if it has
the form $\Theta_{n_q} = \mathrm{diag}_{n_q}(J)$.

\item[(b)] The system variables and the output satisfy the non-demolition condition \cite{BVP91}, \cite{BVP94}:
\begin{equation} \label{xyq_jun14}
\left[x_q(t), ~\left[
    \begin{array}{c}
      y_{q}(r) \\
      y_{q}'(r) \\
    \end{array}
  \right]^T
  \right] =0,~~~ t\geq r\geq 0.
\end{equation}
In other words, the current system variables are compatible with past outputs.

\item[(c)] Define skew-symmetric real matrices $\Theta_w$ and $\Theta_{qq'}$ by means of
\begin{eqnarray}
\left[dw'(t),~   dw'(t)^T\right] &=& 2i \Theta_{w},
  \label{Theta_w}
  \\
\left[\left[
    \begin{array}{c}
      dy_{q}(t) \\
      dy_{q}'(t) \\
    \end{array}
  \right], \left[
    \begin{array}{c}
      dy_{q}(t) \\
      dy_{q}'(t) \\
    \end{array}
  \right]^T
  \right] &=& 2i \Theta_{qq'}, ~~ t\geq 0.
  \label{Theta_y}
 \end{eqnarray}
Then
\begin{equation} \label{ywq_jun14}
\Theta_{qq'} = \left[
    \begin{array}{c}
      D_{q} \\
      D'_{q} \\
    \end{array}
  \right]
 \Theta_w
  \left[
    \begin{array}{c}
      D_{q} \\
      D'_{q} \\
    \end{array}
  \right]^T.
\end{equation}
This simply means that the quantum noise component at the output corresponds to a boson field, just like the input.
\end{description}

It turns out that the when
\begin{equation}
\left[
    \begin{array}{c}
      D_{q} \\
      D'_{q} \\
    \end{array}
  \right] = [I_{2(n_{y_q} + n_{y_q'})} \quad 0_{2(m-n_{y_q} - n_{y_q'})}],
\label{mixed:condition3}
\end{equation}
the above properties are guaranteed when the real constant matrices $A_{qq}$, $B_{q}$, $C_{qq}$, $D_q$, $C'_{qq}$,  and $D'_q$ satisfy the so-called {\it physical realizability
conditions}, \cite[Theorem 3.4]{JNP08}:
\begin{eqnarray}
 B_q \Theta_w \left[\begin{array}{c} D_q \\ D'_q \end{array} \right]^T = - \Theta_{n_q} [\begin{array}{cc} C_{qq} & C'_{qq} \end{array}]^T \label{mixed:condition2}
 \\
A_{qq}\Theta_{n_q}+ \Theta_{n_q} A_{qq} ^{T}+B_{q} \Theta_{w} B_{q}^{T}=0
\label{mixed:condition1}.
\end{eqnarray}
We remark that \eqref{mixed:condition3} is not the most general form possible for $[D_q^T,(D'_q)^T]^T$. The general setting only requires that
\eqref{ywq_jun14} holds without imposing any further conditions on the structure or form of  $[D_q^T,(D'_q)^T]^T$. Moreover,  in the general setting the requirements for properties \textbf{(a)} and \textbf{(b)} are again given by \eqref{mixed:condition2}\footnote{Note that \cite{JNP08} writes \eqref{mixed:condition2} in the equivalent form (that is valid when \eqref{mixed:condition3} holds) $B_{q}\left[
    \begin{array}{c}
      D_{q} \\
      D'_{q} \\
    \end{array}
  \right]^T
= \Theta_{n_q}\left[
    \begin{array}{c}
      C_{qq} \\
      C'_{qq} \\
    \end{array}
  \right]^T\Theta_{w}$} and \eqref{mixed:condition1}, respectively \cite[Theorems 2.1 and 2.2]{NY17}.

\begin{remark}
In fact, as shown in the proof of \cite[Theorem 3.4]{JNP08},   Eq. \eqref{xq_jun14} is equivalent to Eq. \eqref{mixed:condition1}. Moreover, noticing
\[
\left[\left[
    \begin{array}{c}
      dy_{q}(t) \\
      dy_{q}'(t) \\
    \end{array}
  \right], \left[
    \begin{array}{c}
      dy_{q}(t) \\
      dy_{q}'(t) \\
    \end{array}
  \right]^T
  \right] = [dw'(t),~ dw'(t)^T],
\]
 Eq. \eqref{mixed:condition3} leads to  Eq. \eqref{ywq_jun14}. Finally, for the special form of $D_q$ and $D_q'$ as in Eq. \eqref{mixed:condition3},   Eq. \eqref{xyq_jun14} is equivalent to Eq. \eqref{mixed:condition2}.
\end{remark}

Finally, since the system  $G_2$ is a quantum linear system, it has an effective Hamiltonian of the form $H_q=\frac{1}{2}x_q^TR_qx_q+x_q^TK_q u(t)$, where $R_q =R_q^T  \in \mathbb{R}^{2n_q \times 2n_q}$,  and $K_q=-\Theta_{n_q}E  \in \mathbb{R}^{2n_q \times 2m}$.  The first term of $H_q$, namely  $\frac{1}{2}x_q^TR_qx_q$, is the isolated Hamiltonian (also called free Hamiltonian) of the system $G_2$, while its second term $x_q^TK_q u(t)$, often called a control Hamiltonian, is induced by the coupling to the external environment through the classical signal $u(t)$ which is a vector of real locally square integrable functions. Some discussions on effective Hamiltonians for the linear case can be found in, e.g., \cite[Chapter 6]{WM10} (in particular Eq. (6.219) in \cite{WM10}), and discussions on more general nonlinear case can be found in, e.g., \cite[Eq. (2) and Sec. 5]{KC08} and \cite[Sec. 1-C]{TV09}. More details on the implementation of $x_q^TK_qu$ using classical devices can be found in,  e.g., \cite{YNJP08}, \cite{Gough08}.

\subsection{Mixed quantum-classical linear stochastic systems with quantum inputs and quantum outputs}\label{subsec:intro-mixed}

Equations \eqref{quantum-eq1}-\eqref{quantum-eq3}  look superficially like the classical state space
equations  familiar to control engineers, but in fact are fundamentally
different because they are equations for a collection of quantum degrees of freedom (noncommutative variables), not a collection of classical degrees of freedom (commutative variables). Even so,  the classical system described by \eqref{classical-eq1}-\eqref{classical-eq4} and the quantum system given by \eqref{quantum-eq1}-\eqref{quantum-eq3} may be interconnected as a mixed quantum-classical  system via appropriate interfaces (homodyne detectors and modulators\footnote{Homodyne detectors are used to measure  quadratures of an optical field and the measurement
outputs are classical signals (photo-current) that can be injected into classical systems. Modulators are utilized to merge quantum and classical signals to form a third signal with desirable characteristics of both in a manner suitable for transmission to quantum systems.}\footnote{As introduced in Subsection \ref{sec:preliminaries-probability}, the measurement results may be seen as  operation of selecting classical elements of quantum signals while the modulation results may be viewed as quantum representation of classical signals \cite{WNZJ13}.}). As introduced in Subsection \ref{sec:preliminaries-probability}, any classical probability model ($\Omega, F, \nu$) can be
viewed as a commutative quantum probability model ($\mathcal{A}, \rho$), which means that classical components can be treated within the formalisms of quantum mechanics by embedding them as commutative subsystem in a quantum system. The problem of putting quantum and classical degrees of freedom within the same formalism has also been discussed in \cite{NK2001,SP07,JNP08,GJ09a, Nurdin2011,TC12} and the references therein. In particular, in the physics literature the formalism is known as the Koopman-von Neumann formulation (of classical mechanics) and the embedding of a classical dynamical system in a quantum one is referred to as a ``quantum mechanics-free subsystem'' \cite{TC12}. In this paper, we aim to  develop a mathematical representation for a class of mixed quantum-classical linear stochastic systems.

We briefly review some  results about mixed quantum-classical linear stochastic systems with quantum inputs and quantum outputs studied in \cite{JNP08} and \cite{Nurdin2011}. Let $x$ have both quantum and classical degrees of freedom, such
that $x=[x_q^T \quad x_c^T]^T$. To be interpreted a classical variable, we require that the entries of ${x}_c(t)$
commute with one another and with entries of the vector of quantum observables $x_q(t)$. Thus, the commutation relation for $x(t)$
satisfies
\[
 [x(t), ~ x(t)^T] = 2 i \Theta_n,
 \]
  where
$\Theta_n=\mathrm{diag}(\Theta_{n_q},0_{n_c\times
n_c})$. In particular, if $\Theta_{n_q}=\mathrm{diag}_{n_q}(J)$, then $\Theta_n$
is said to be {\em degenerate canonical},
\cite{JNP08}. Actually, we require more, that $x_c(t)$ is isomorphic to a classical stochastic process. That is, $[x_c(t),x_c(s)]^T=0$ for all $s \geq 0$ not necessarily equal to $t$. This does in fact hold, and we will say more about this immediately after Theorem \ref{theorem-1}. Following \cite{JNP08},    we thus consider a linear mixed quantum-classical system of the form
\begin{eqnarray}
dx(t)&=&Ax(t)dt+Bdw(t), \label{standard-form1_a}\\
dy_q(t)&=&C_qx(t)dt +D_qdw(t), \label{standard-form1_b}
\end{eqnarray}
where $w(t)$ and $y_q(t)$ are quantum input and output fields, respectively, and  $A \in \mathbb{R}^{ n
\times n}$,
 $B \in \mathbb{R}^{n \times 2m}$, $C_q \in \mathbb{R}^{2n_{y_q}
\times n}$ and $D_q\in \mathbb{R}^{2n_{y_q}\times 2m}$, $n=2n_q+n_c$.  As discussed in Subsection \ref{sec:preliminaries-probability}, if we
are given a component of a vector of classical system variables
$x_c$ denoted by $x_{c_k}$, we may consider $x_{c_k}$ as one of the
quadratures of a quantum harmonic oscillator, say the amplitude
quadrature $q_k$. Then we may define an {\it augmentation} of $x_{c_k}(t)$, say
$\tilde{x}_k(t) = \left[
\begin{array}{c}
q_k(t) \\
p_k(t)\\
\end{array}
\right]
$. Therefore, $x(t)$
can be embedded in a larger vector
$\tilde{x}(t) = [x(t)^T\quad
\eta(t)^T ]^T$, where any element of
$\eta(t) = [p_1(t)~ p_2(t)~
\cdots~ p_{n_c}(t)]^T$ commutes with any component of
$x_q(t)$, and are conjugate to
the components of $x_c(t)$, satisfying
$[x_{c_j}(t),
\eta_k(t)]=2i\delta_{jk}$, where $\delta_{jk}$ is the Kronecker
delta function. As a result, the commutation relation for
$\tilde{x}(t)$ is $[\tilde{x}(t),~ \tilde{x}(t)^T] = 2 i \tilde{\Theta}$, where
\[
\tilde{\Theta}\!=\!\!\left[\!
\begin{array}{cc}
\Theta_n &
\left[
\begin{array}{c}
0 \\
I_{n_c} \\
\end{array}
\right]
\\
\left[
\begin{array}{cc}
0& -I_{n_c}\\
\end{array}
\right]
& 0 \\
\end{array}
\!\right]
\] is an invertible matrix satisfying $\tilde{\Theta}\tilde{\Theta}=-I$ and $\tilde{\Theta}=-\tilde{\Theta}^T$.   Moreover, as shown in \cite{JNP08}, there is an  augmentation of the system \eqref{standard-form1_a}-\eqref{standard-form1_b}  in terms of
$\tilde{x}$, which  can be written as
\begin{eqnarray}
d\tilde{x}(t)&=&\tilde{A}\tilde{x}(t)dt+\tilde{B}dw(t)
\label{form4},
\\
dy_q(t)&=&\tilde{C}\tilde{x}(t)dt+D_q dw(t)
\label{form5},
\end{eqnarray}where $\tilde{A}\!\!=\!\!\left[
\begin{array}{cc}
A & 0 \\
A' &A'' \\
\end{array}
\right]$, $\tilde{B}\!=\!\!\left[
\begin{array}{c}
B \\
B' \\
\end{array}
\right]$, and $\tilde{C}\!=\!\!\left[
\begin{array}{cc}
C_{q}\quad 0 \\
\end{array}
\right]$.

We first have the following definition when $D_q$ takes on a  particular form.

\begin{definition}\label{defn:mixed-qc-0} [\cite{JNP08}]
Let $D_q=I_{2m}$ or  $D_q=[I_{2n_{y_q}} \quad 0_{2(m-n_{y_q})}]$.  The mixed quantum-classical system \eqref{standard-form1_a}-\eqref{standard-form1_b} with quantum inputs and quantum outputs is
physically realizable if there exists an augmentation of the form \eqref{form4} and  \eqref{form5} that is a physically realizable fully quantum system. That is, if there exist matrices $A'$, $A''$, $B'$ such that \eqref{mixed:condition2}-\eqref{mixed:condition1} hold with matrices
$A_{qq}$, $B_{q}$,
$\left[
    \begin{array}{c}
      C_{qq} \\
      C'_{qq} \\
    \end{array}
  \right]
$ and  $\left[
    \begin{array}{c}
      D_{q} \\
      D'_{q} \\
    \end{array}
  \right]$
replaced by corresponding matrices $\tilde{A}$, $\tilde{B}$, $\tilde{C}$, $D_q$, respectively.
\end{definition}

The following result gives the physical realizability conditions for the mixed quantum-classical system \eqref{standard-form1_a}-\eqref{standard-form1_b}  that follow from those for the fully quantum system \eqref{form4}-\eqref{form5}. Note that the conditions below do not depend on $A'$, $A''$, $B'$. That is, the conditions are intrinsic on the system \eqref{standard-form1_a}-\eqref{standard-form1_b}. If these conditions are fulfilled then there exist suitable choices of $A'$, $A''$, $B'$ to construct a physically realizable augmentation.

\begin{theorem}\label{theorem-1} [\cite{JNP08}]
Let $D_q=I_{2m}$ or  $D_q=[I_{2n_{y_q}} \quad 0_{2(m-n_{y_q})}]$. The mixed quantum-classical system \eqref{standard-form1_a}-\eqref{standard-form1_b} with quantum inputs and quantum outputs is
physically realizable if and only if
\begin{eqnarray*}
A \Theta_n + \Theta_n A^T + B {\rm diag}_{m}(J)B^T = 0,\\
\Theta_{n} C^T = -B {\rm diag}_{m}(J) D_q^T.
\end{eqnarray*}
\end{theorem}

From the above theorem, it will be guaranteed that $[x(t),x(t)^T]=2i \Theta_n$ for all $t \geq 0$. However, as alluded to earlier, for $x_c(t)$ to be interpretable as a classical stochastic process, we require that $[x_c(t),x_c(s)^T]= 0$ for all $s \geq 0$ not necessarily equal to $t$.  In a more general setting to be given in Definition \ref{dfn:realize1}, we make this explicit by imposing the requirement that the entries of $x_c(t)$ commute will every entry of $x_c(s)$ for any time $s \geq 0$ not necessarily equal to $t$. That this is indeed the case will be seen in the proof of Theorem \ref{Them-form} where it emerges as an easy consequence of the equal time commutation relations $[x_c(t),x(t)^T]=0$. In other words, the equal time commutation relations that forms a basis for the physical realizability of the augmentation \eqref{form4} and  \eqref{form5} is enough to characterize the mixed quantum-classical system given by \eqref{standard-form1_a}-\eqref{standard-form1_b}. Yet another way to view this is that the definition of physical realizability in Definition \ref{defn:mixed-qc-0}  given through an augmentation of \eqref{standard-form1_a}-\eqref{standard-form1_b} is perfectly consistent with $x_c(t)$ being isomorphic to a classical stochastic process, despite the fact that this requirement is not explicitly stated in the definition.

\begin{remark}\label{rem:jun25}
We have the following observations for the abstractly defined mixed quantum-classical linear stochastic system \eqref{standard-form1_a}-\eqref{standard-form1_b}  with quantum inputs and outputs, as studied in \cite{JNP08}:
\begin{description}
\item[(a)] The inputs and outputs are all purely quantum;

\item[(b)]  The matrix $D_q$ is in the form of $D_q=I_{2m}$ (or  $D_q=[I_{2n_{y_q}} \quad 0_{2(m-n_{y_q})}]$ if $n_{y_q}<m$).

\item[(c)]  It is not immediately apparent how quantum and  the classical components are interconnected and what are the interfaces that are required  make the interconnection.
\end{description}
In the sections that will follow, we will relax the requirements \textbf{(a)} and \textbf{(b)} and also address \textbf{(c)} in a more general setting.
\end{remark}

\section{Canonical representation of  mixed linear stochastic systems}\label{sec:mod}
In this section, we give two forms  for mixed quantum-classical linear stochastic systems described by LSDEs, one being a  {\it general} form  in which the mixed system is often obtained  in real experiments and the other being a  {\it standard} form in which the mixed system can be easily decomposed for analysis and synthesis.  We also derive relations between the two forms. Notice that in this paper we  allow the general form to include classical inputs and outputs as well as scattering processes,  which are more general than  the mixed quantum-classical linear stochastic systems of the form \eqref{standard-form1_a}-\eqref{standard-form1_b}  with quantum inputs and outputs, as discussed in Subsection \ref{subsec:intro-mixed}; cf. Remark \ref{rem:jun25}.

\subsection{A standard form for mixed linear stochastic systems with quantum inputs and mixed outputs}\label{s-form}

Consider the following mixed linear stochastic system with quantum inputs and mixed outputs:
\begin{eqnarray}
dx(t)&=&Ax(t)dt+Bdw(t), \label{standard-form_a}\\
dy(t)&=&Cx(t)dt +Ddw(t). \label{standard-form_b}\
\end{eqnarray}
As specified before, the system variables are $x=[x_q^T \quad x_c^T]^T$, the system outputs are $y=[y_q^T\quad y_c^T]^T$. Define a constant real matrix $F_y$ by
 \begin{equation} \label{eq:Fy}
dy(t)dy(t)^T = F_ydt.
 \end{equation}
 Also, define a real skew-symmetric matrix $\Theta_{y_q}$ in terms of
 \begin{equation} \label{CCR:yq}
  [dy_q(t),~ dy_q(t)^T]= 2i\Theta_{y_q}.
 \end{equation}
Clearly,  the mixed output $y_q(t)$ is a vector of dimension $n_{y}=2n_{y_q}+n_{y_c}$.  For later use, the system input $w(t)$ is partitioned to be  $w(t)=[w_1(t)^T\quad w_2(t)^T]^T$ where $w_1(t)$ is of dimension $2n_{w_1}$ and $w_2(t)$ is of dimension $2n_{w_2}$.  However, instead of being of the form $I$ or $[\begin{array}{cc} I & 0 \end{array}]$ as in Eq. \eqref{mixed:condition3}, or equivalently $D_q$ specified in Theorem \ref{theorem-1}, in general, the matrix $D$  is associated with  gauge processes representing the photon exchange among the external fields represented here by $w(t)$.

\begin{remark}
It will be shown later in Remark \ref{rem:jun25_ww} that
\begin{equation}
[dw(t), ~dw(t)^T] =  2i\Theta_w,
\end{equation}
where the skew-symmetric matrix $\Theta_w$  is given in Eq. \eqref{Theta_w}.
\end{remark}

The transfer function for
system \eqref{standard-form_a}-\eqref{standard-form_b} is
\begin{eqnarray*}
\Xi_{S}(s)=\left[\begin{tabular}{l|ll}
$A$ & $B$\\
\hline \vspace{-2em}\\
$C$ & $D$\\
\end{tabular}\right](s)=C\left(sI_{n}-A\right)^{-1}B+D.
\end{eqnarray*}

\begin{definition}\label{defn:standard-form}
The mixed quantum-classical linear stochastic system
 \eqref{standard-form_a}-\eqref{standard-form_b} is said to be standard  if the following
 holds:
\begin{enumerate}
\item $\Theta_n=\mathrm{diag}(\Theta_{n_q},0_{n_c\times n_c})$ with $\Theta_{n_q}=\mathrm{diag}_{n_q}(J)$.
\item $\Theta _{w}=\mathrm{diag}_m(J)$.
\item The matrix $F_y$ defined in Eq. \eqref{eq:Fy} satisfies $F_{y}=I_{n_{y}}+i\, \mathrm{diag}(\Theta_{y_q},
0_{n_{{y_c}}\times n_{{y_c}}})$, where $\Theta_{y_q}=\mathrm{diag}_{n_{y_q}}(J)$.
\end{enumerate}
\end{definition}

Now let the matrices $A$, $B$,  $C$, $D$ be partitioned compatibly with
partitioning of
$x(t)$ into
$x_q(t)$ and $x_c(t)$ as
\begin{equation}\label{ABCD}
A=\left[
\begin{array}{cc}
A_{qq} & A_{qc} \\
A_{cq} & A_{cc} \\
\end{array}
\right]
, \quad B=\left[
\begin{array}{c}
B_q \\
B_c \\
\end{array}
\right], \quad C\!=\left[
\begin{array}{c}
C_{q} \\
C_{c}\\
\end{array}\right]\!=\!\left[
\begin{array}{cc}
C_{qq}& C_{qc} \\
C_{cq} &C_{cc} \\
\end{array}
\right], \quad D=\left[
\begin{array}{c}
D_{q} \\
D_{c}\\
\end{array}
\right].
\end{equation}
 Then, the system \eqref{standard-form_a}-\eqref{standard-form_b} can be rewritten as
follows:
\begin{eqnarray}
dx_q(t)&=&[
\begin{array}{cc}
A_{qq} \quad A_{qc} \\
\end{array}
]x(t)dt+B_qdw(t),\label{form1}\\
dx_c(t)&=&[
\begin{array}{cc}
A_{cq}\quad A_{cc} \\
\end{array}
]x(t)dt+B_cdw(t),\label{form1.2}\\
dy_q(t)&=&[
\begin{array}{cc}
C_{qq} \quad C_{qc} \\
\end{array}
]x(t)dt+D_qdw(t),\label{form2}\\
dy_c(t)&=&[
\begin{array}{cc}
C_{cq}\quad C_{cc} \\
\end{array}
]x(t)dt+D_cdw(t).
\label{form3}
\end{eqnarray}

\begin{remark}
The features presented in Definition \ref{defn:standard-form}  allow us to consider
classical variables $x_c(t)$, characterized by zero commutation relations, as
well as classical noise processes $y_c(t)$, corresponding to the absence of
the imaginary part in the Ito products, \cite{JNP08},
\cite{Nurdin2011}. The first item of Definition \ref{defn:standard-form} indicates that $x(t)$ has both quantum and classical degrees of
freedom, where $\Theta_{n_q}$ corresponds to the quantum degrees of
freedom $x_q(t)$, while $0_{n_c\times n_c}$ corresponds to the classical
degrees of freedom $x_c(t)$. The second item of Definition
\ref{defn:standard-form} shows that input signals of the system
\eqref{standard-form_a}-\eqref{standard-form_b} are fully quantum. Finally, let
\[
\Theta_{y} =\frac{[dy(t),~ dy(t)^T]}{2i}=\frac{F_y - (F_y)^T}{2i}.
\]
Clearly,
\[
\Theta_{y}=\mathrm{diag}(\Theta_{y_q},
0_{n_{y_c}}).
\]
Therefore, the third item of
Definition \ref{defn:standard-form} implies that $\Theta_{y_q}$
corresponds to quantum outputs $y_q(t)$ while the matrix
$0_{n_{y_c}}$
corresponds to classical outputs $y_c(t)$. Finally, in analogy to Eq. \eqref{ywq_jun14}, we have
\begin{eqnarray}\label{ccr-output1}
\Theta_{y}=D\Theta_{w}D^T.
\end{eqnarray}

\end{remark}

\begin{remark}The difference between the
mixed linear systems \eqref{standard-form1_a}-\eqref{standard-form1_b} and
\eqref{standard-form_a}-\eqref{standard-form_b} is that the latter explicitly exhibits
classical output signals, and the matrix $D$ has a more general form
satisfying condition \eqref{ccr-output1}, which is equivalent to the
following equations:
\begin{eqnarray}
D_{q}\Theta_{w} D^T_{q}&=&\Theta_{y_q}\label{constraint8},\\
D_{q}\Theta_{w}D^T_{c}&=&0\label{constraint9},\\
D_{c}\Theta_{w}D^T_{c}&=&0\label{constraint10}.
\end{eqnarray}
\end{remark}

\subsection{A general form for mixed linear stochastic systems with mixed inputs and mixed outputs}\label{g-form}

In Definition \ref{defn:standard-form},  the quantum-classical nature of the   standard  form is captured in the  matrices  $\Theta_n$, $\Theta _{w}$, $F_{y}$   specifying the commutation relations
of the system and signal. In general, we
may take the commutation matrix  to be an arbitrary  real skew-symmetric matrix, while
the Ito matrix $F$ is a free non-negative Hermitian matrix.  To this end, consider a general form for linear mixed quantum-classical
stochastic systems given by
\begin{eqnarray}
d\mathbf{x}(t)&=&\mathbf{A}\mathbf{x}(t)dt+\mathbf{B}dv(t),\label{general-form_a}\\
d\mathbf{y}(t)&=&\mathbf{C}\mathbf{x}(t)dt+\mathbf{D}dv(t), \label{general-form_b}
\end{eqnarray}where $\mathbf{A} \in \mathbb{R}^{n
\times n}$,
$\mathbf{B} \in \mathbb{R}^{n \times n_v}$, $\mathbf{C} \in \mathbb{R}^{n_{\mathbf{y}}
\times n}$ and $\mathbf{D}\in \mathbb{R}^{n_{\mathbf{y}}\times n_v}$;
$\mathbf{x}(t)$ includes quantum and classical system variables satisfying
the commutation relation, such that
$[\mathbf{x}(t),~~ (\mathbf{x}(t))^T] = 2 i \hat{\Theta}_n$ with
a skew-symmetric matrix
$\hat{\Theta}_n$; the vector $v(t)$ represents the
input signals, which contains quantum and classical noises;
$\mathbf{y}(t)$ represents mixed quantum-classical outputs.
$F_v$ and $F_\mathbf{y}$ are
nonnegative definite Hermitian matrices satisfying
$dv(t) dv(t)^T = F_{v}dt$ and
$d\mathbf{y}(t) d\mathbf{y}(t)^T = F_{\mathbf{y}}dt$.  Define
\[
\Theta_{\bf y} = \frac{F_{\bf y} - (F_{\bf y})^T}{2i}.
\] The transfer function $\Xi_G(s)$
for a system of the form \eqref{general-form_a}-\eqref{general-form_b} is given by
\begin{eqnarray*}
\Xi_{G}(s)=\left[\begin{tabular}{l|ll}
$\mathbf{A}$ & $\mathbf{B}$\\
\hline \vspace{-2em}\\
$\mathbf{C}$ & $\mathbf{D}$\\
\end{tabular}\right](s)=\mathbf{C}\left(sI_{n}-\mathbf{A}\right)^{-1}\mathbf{B}+\mathbf{D}.
\end{eqnarray*}

\subsection{Relations between the General and Standard Forms}
The  standard  form  \eqref{standard-form_a}-\eqref{standard-form_b} and the general form \eqref{general-form_a}-\eqref{general-form_b}
can be related by the following lemmas and
theorem:

\begin{lemma}\label{lemma1}
Given an arbitrary
$n\times n$ real skew-symmetric matrix
$\hat{\Theta}_n$ ($n\geq2$),
there exists a real nonsingular matrix
$P_n$ and a block diagonal
matrix
$\Theta_n=\mathrm{diag}(\Theta_{n_q},0_{n_c\times
n_c})$ such that
\begin{eqnarray}
\Theta_n=P_n\hat{\Theta}_n P_n^{T}.
\end{eqnarray}
\end{lemma}
A similar proof of Lemma \ref{lemma1} can be found in
\cite[Theorem 2.5.8]{HJ1985} and hence the proof is omitted here.

\begin{lemma}\label{lemma2}
Given an arbitrary $m \times m$ nonnegative definite Hermitian
matrix $F_v$, there exists a $2m \times 2m$ matrix
$F_{w}$=$I_{2m}$+$i\ \mathrm{diag}_m(J)$ and a $m \times 2m$ real
matrix $W$ such that
\begin{eqnarray}\label{W1}
F_v=WF_{w}W^{T}.
\end{eqnarray}
\end{lemma}

\quad \quad $\emph{Proof}$:
Hermitian matrices $F_v$ and $F_{w}$ can be diagonalized by unitary
matrices
$U_v$ and $U_w$, respectively,
such that \begin{eqnarray}
F_v&=&U_v\Lambda_v U_v^{\dagger} \label{F1},\\
F_{w}&=&U_w\Lambda_{w}U_w^{\dagger}, \label{F2}
\end{eqnarray}where
$\Lambda_v$=diag$(\lambda_1,
\lambda_2,\cdots \lambda_m)$, ($\lambda_j\geq 0$ is an eigenvector
of $F_v$), $\Lambda_{w}$=diag$_m\left(\left[
\begin{array}{cc}
0, & 0 \\
0, & 2 \\
\end{array}
\right]\right)$, $U_w$=diag$_m\left(\frac{\sqrt{2}}{2}\left[
\begin{array}{cc}
i ,& i \\
-1, & 1 \\
\end{array}
\right]\right)$. Since $\Lambda_v$ and
$\Lambda_{w}$ are two real diagonal matrices, there exists a
$m\times 2m$ complex matrix
$Q=[q_1, q_2, \cdots, q_{2m}]$
such that
\begin{eqnarray}\label{F3}
\Lambda_v=Q\Lambda_{w}Q^{\dagger}.
\end{eqnarray}In order to let \eqref{F3} hold, for simplicity we choose
{\setlength\arraycolsep{1.5pt}$q_2=\left[
\begin{array}{cccc}
\sqrt{\frac{\lambda_1}{2}} & 0 & \cdots & 0 \\
\end{array}
\right]^T$
, $q_4=\left[
\begin{array}{ccccc}
0 & \sqrt{\frac{\lambda_2}{2}} & 0&\cdots & 0 \\
\end{array}
\right]^T$, $\cdots$, $q_{2m}=\left[
\begin{array}{cccc}
0 & \cdots & 0 &\sqrt{\frac{\lambda_{m}}{2}}
\end{array}
\right]^T$}, and $q_1, q_3, \cdots,
q_{2m-1}$ now are arbitrary column vectors of length $m$ and to be determined later. Combining
\eqref{F1}, \eqref{F2} and
\eqref{F3} gives
\begin{eqnarray}\label{W2}
F_v=U_vQ U_w^{\dagger}F_{w}(UQ U_w^{\dagger})^{\dagger}.
\end{eqnarray}
Let $W$ be defined as $W=U_vQ U_w^{\dagger}$. Then we have
\begin{eqnarray}\label{W3}
U_vQ=[U_vq_1, U_vq_2, \cdots, U_vq_{2m}].
\end{eqnarray}Next, we will show
that $Q$ can be chosen to let $W$ be real. Observing the structure
of $U_w$, such that
$$U_w=\mathrm{diag}_m\left(\frac{\sqrt{2}}{2}\left[
\begin{array}{cc}
i ,& i \\
-1, & 1 \\
\end{array}
\right]\right),$$ we require that $q_1, q_3, \cdots,
q_{2m-1}$ be chosen as
\begin{eqnarray*}
q_1=-U_v^{\dagger}U_v^{\#}q_2,\quad
q_3=-U_v^{\dagger}U_v^{\#}q_4,
\quad  \cdots,  \quad
q_{2m-1}&=&-U_v^{\dagger}U_v^{\#}q_{2m}.
\end{eqnarray*}
The matrix $Q$ is hence constructed as
\begin{eqnarray*}\label{F5}
Q=\left[
\begin{array}{ccccccc}
-U_v^{\dagger}U_v^{\#}q_2,& q_2,& -U_v^{\dagger}U_v^{\#}q_4,& q_4,&
\cdots& -U_v^{\dagger}U_v^{\#}q_{2m},& q_{2m}\\
\end{array}
\right].
\end{eqnarray*}

We can get the representation \eqref{W1} with
$W=U_vQ U_w^{\dagger}$.
\hfill $\blacksquare$

Let us look at an example applying Lemma \ref{lemma2}.

$\mathbf{Example \  1}$: Consider a nonnegative definite Hermitian
matrix given by
$$F_v\!=\!\!\left[\!\!\!
              \begin{array}{ccc}
 8.9286 & -0.2143 + 4.8107i  & 0.1429 + 7.2161i\\
  -0.2143 - 4.8107i &  8.3571 + 0.0000i  & 0.4286 - 2.4054i\\
   0.1429 - 7.2161i  & 0.4286 + 2.4054i &  8.7143  \\
              \end{array}
            \!\!\!\right].$$ It is easily obtained that $F_v=U_v\Lambda_v U^{\dagger}$ with
            $$U_v=\left[\!\!
                                \begin{array}{ccc}
   0.6814   &          0.6814  &           0.2673 \\
  -0.1572 - 0.3922i & -0.1572 + 0.3922i &  0.8018\\
   0.1048 - 0.5883i  & 0.1048 + 0.5883i & -0.5345\\
                                \end{array}
                              \right]$$ and
                              $$\Lambda_v=\left[
                                                      \begin{array}{ccc}
                                                        18 & 0& 0 \\
                                                        0 & 0 & 0 \\
                                                        0& 0 & 8 \\
                                                      \end{array}
                                                    \right].$$
  Now following the construction in the proof of Lemma \ref{lemma2}, we want to find a real matrix $W$. Choosing $q_2=[3\quad 0 \quad 0]^T, q_4=[0\quad  0 \quad 0 ]^T$ and $q_6=[0\quad  0\quad  2]^T$ we get $q_1=[0\quad  -3 \quad 0]^T, q_3=[0 \quad 0 \quad 0 ]^T$ and $q_5=[0\quad  0 \quad -2]^T$. So the matrix $Q=\left[
                                                                                             \begin{array}{cccccc}
                                                                                               0 & 3 & 0 & 0 & 0 & 0 \\
                                                                                               -3 & 0 & 0 & 0 & 0 & 0 \\
                                                                                               0 & 0& 0 & 0& -2 & 2 \\
                                                                                             \end{array}
                                                                                           \right]$. It follows from the above construction that  $W=\left[
                                                                                                                     \begin{array}{cccccc}
                                                                                                                       0 &  2.8909 & 0 &0 & 0 & 0.7559\\
                                                                                                                       -1.6641 &  -0.6671 &  0 &  0 &  0 &  2.2678 \\
                                                                                                                       -2.4962  &  0.4447 & 0 & 0 & 0 & -1.5119 \\
                                                                                                                     \end{array}
                                                                                                                   \right]$. It is easily checked that $F_v=WF_{w}W^{T}$ with $F_{w}$=$I_{6}$+$i\ \mathrm{diag}_3(J)$.

\begin{theorem}\label{two-relation}
Given a mixed quantum-classical stochastic system of the general
form  \eqref{general-form_a}-\eqref{general-form_b}, there exists a corresponding  standard  form  \eqref{standard-form_a}-\eqref{standard-form_b}.
\end{theorem}
\quad \quad $\emph{Proof}$: By Lemmas \ref{lemma1} and \ref{lemma2}, there exist matrices
$P_n$, $W$ and
$P_\mathbf{y}$ such that the coordinate transformations
\[
x = P_n {\bf x}, \quad y=P_\mathbf{y}\mathbf{y}, \quad w = W^T v
\]
yields
\begin{eqnarray}\label{relations}
\left.
\begin{array}{ccc}
\Theta_n=P_n\hat{\Theta}_n P_n^T,\quad \Theta_y=P_\mathbf{y}\Theta_\mathbf{y}P_\mathbf{y}^T,\quad \Theta_v=W\Theta_{w}W^{T},
\\
\quad A=P_n\mathbf{A}P_n^{-1}, \quad
B=P_n\mathbf{B}W, \quad C=P_\mathbf{y}\mathbf{C}P_n^{-1},\quad D=P_\mathbf{y}\mathbf{D}W.
\\
\end{array}
\right\}
\end{eqnarray}
Substituting  \eqref{relations} into \eqref{general-form_a}-\eqref{general-form_b}
gives  \eqref{standard-form_a}-\eqref{standard-form_b}. Now, we can verify the following
relation between the standard  $\Xi_S(s)$ and general
$\Xi_G(s)$ transfer functions:
\begin{eqnarray*}
\Xi_{S}(s)&=&C\left(sI_{n}-A\right)^{-1}B+D\\
&=&P_\mathbf{y}\mathbf{C}P_n^{-1}\left(sP_nP_n^{-1}-P_n\mathbf{A}P_n^{-1}\right)^{-1}P_n\mathbf{B}W+P_\mathbf{y}\mathbf{D}W\\
&=&P_\mathbf{y}\left(\mathbf{C}\left(sI_{n}-\mathbf{A}\right)^{-1}\mathbf{B}+\mathbf{D}\right)W\\
&=&P_\mathbf{y}\Xi_{G}(s)W.\vspace{-1em}
\end{eqnarray*}
Thus, the general form \eqref{general-form_a}-\eqref{general-form_b} can be linearly
transformed into its corresponding   standard  form
 \eqref{standard-form_a}-\eqref{standard-form_b}.
\hfill  $\blacksquare$

\section{Physical realizability of mixed quantum-classical linear stochastic systems} \label{sec:realization}

In this section, we will introduce the definition of physical
realizability of the   standard  form \eqref{standard-form_a}-\eqref{standard-form_b} and a theorem on necessary
and sufficient conditions for its  physical realizability. Analogous
physical realizability definition and conditions for the general
form \eqref{general-form_a}-\eqref{general-form_b} are also presented in this section.

\subsection{Physical realizability for the standard form}

The following concepts and lemmas will be used for introducing the definition of physical realizability of the system  \eqref{standard-form_a}-\eqref{standard-form_b}.

The Belavkin's nondemolition principle requires  an  observable $X(t)$  at a time instant $t$ to be compatible with the past
output process $Y(s)$ ($s\leq t$) \cite{BVP91}, \cite{BVP94}, that is:
\begin{eqnarray}\label{non-demolition}
[X(t), ~Y(s)^T]=0, \  \forall\ t\geq s \geq 0.
\end{eqnarray}

Condition \eqref{non-demolition} is known as  {\em  non-demolition condition}.

\begin{lemma}\label{nondemolition0}
Non-demolition condition
$[\tilde{x}(t), ~y_q (s) ^T]=0$,
$\forall$ $t\geq s \geq 0$ for the   augmented  system
\eqref{form4}-\eqref{form5} of the system  \eqref{standard-form1_a}-\eqref{standard-form1_b}
holds, if and only if
\begin{eqnarray}\label{new-conditionBCD}
\tilde{B}\Theta_{w}D_q^T
=-\tilde{\Theta}
\tilde{C}^T.
\end{eqnarray}
\end{lemma}

\quad \quad $\emph{Proof}$:
First, we will argue that
$[\tilde{x}(t), y_q(s)^T]=0$ is
equivalent to $[\tilde{x}(t), y_q^T(t)]=0$, for all $t\geq s \geq
0$. Let
$g_s(t)=[\tilde{x}(t),
y_q(s)^T]$, for all $t\geq s \geq 0$, where $s$ is fixed. From
$[\tilde{x}(t), y_q(t)^T]=0$
for all $t\geq s \geq 0$, we can infer that $g_s(s)=0$ and then
have \begin{eqnarray*}
dg_s(t)&=&d[\tilde{x}(t), y_q(s)^T]\\
&=&[d\tilde{x}(t), y_q(s)^T]\\&=&\tilde{A}[\tilde{x}(t),
y_q(s)^T]dt\\
&=&\tilde{A}g_s(t)dt.
\end{eqnarray*}
Solving the above equation gives $g_s(t)=\mathrm{exp}\left(\tilde{A}(t-s)\right)g_s(s)=0$. Therefore, $[\tilde{x}(t),
y_q(t)^T]=0$ implies
$[\tilde{x}(t), y_q(s)^T]=0$,
for all $t\geq s \geq 0$.  Conversely, it is  trivial to
verify that $[\tilde{x}(t), y_q(s)^T]=0$ for all $t\geq s \geq 0$
implies $[\tilde{x}(t), y_q(t)^T]=0$ for all $t \geq 0$.

Thus, we just need to consider the case where $t=s$. Let
$g(t)=[\tilde{x}(t), y_q (t) ^T]$ with $g(0)=0$ and then we have
\begin{eqnarray*}
dg(t)&=&d[\tilde{x}(t), y_q (t) ^T]\\
&=&[d\tilde{x}(t), y_q (t) ^T]+[\tilde{x}(t), dy_q (t) ^T]+[d\tilde{x}(t), dy_q (t) ^T]\\
&=&\tilde{A}g(t)dt+2i(\tilde{\Theta}\tilde{C}^T+\tilde{B}\Theta_{w}D_q^T)dt.
\end{eqnarray*}Solving the above equation gives
\begin{eqnarray}\label{proof-eq1}
g(t)=\mathrm{exp}(\tilde{A}t)g(0)+2i\int^{t}_{0}\mathrm{exp}(\tilde{A}(t-\tau))\left(\tilde{\Theta}\tilde{C}^T+\tilde{B}\Theta_{w}D_q^T\right)d\tau.
\end{eqnarray} It can be easily verified from \eqref{proof-eq1} that $g(t)=0$ holds for all $t\geq 0$, if
and only if $
\tilde{\Theta}\tilde{C}^T+\tilde{B}\Theta_{w}D_q^T=0$, which is Eq. \eqref{new-conditionBCD}.
\hfill  $\blacksquare$

\begin{lemma}\label{nondemolition}
Non-demolition condition $[x(t), y (s) ^T]=0$, $\forall$ $t\geq s
\geq 0$ for the system  \eqref{standard-form_a}-\eqref{standard-form_b} holds, if and only if
\begin{eqnarray}\label{newconditionBCD1}
B\Theta_{w}D^T
=-\Theta_n
C^T.
\end{eqnarray}
\end{lemma}

The proof of Lemma \ref{nondemolition} is similar to that of Lemma
\ref{nondemolition0} and is thus  omitted.

For a better understanding of Definitions \ref{pre-definition} and
\ref{dfn:realize1} to be given later, a discussion regarding the physical
realizability of  the  standard  form
 \eqref{standard-form_a}-\eqref{standard-form_b} will be given   first. The system
\eqref{standard-form_a}-\eqref{standard-form_b} can be divided into two parts: one is the
system \eqref{standard-form1_a}-\eqref{standard-form1_b}, or equivalently the system
(\ref{form1})-(\ref{form2}),  with $D_q$ satisfying
Eq. \eqref{constraint8}; the other is the output equation
(\ref{form3}). Therefore, the system \eqref{standard-form_a}-\eqref{standard-form_b} is physically
realizable if the two parts are both physically realizable. First,
we consider physical realizability conditions of the system
\eqref{standard-form1_a}-\eqref{standard-form1_b}. From the structure of system matrices of
the  augmented  system \eqref{form4}-\eqref{form5}, it is clear
that the dynamics of $x(t)$ of system \eqref{standard-form1_a}-\eqref{standard-form1_b} embedded in system \eqref{form4}-\eqref{form5} are not affected by
the augmentation, and moreover, it will be shown in the
proof of Theorem \ref{Them-form} below that  matrices $A', A'', B'$ in system
\eqref{form4}-\eqref{form5} can be chosen to preserve commutation
relations for augmented system variables $\tilde{x}$ .
As given in Definition \ref{defn:mixed-qc-0}, the system \eqref{standard-form1_a}-\eqref{standard-form1_b}  with
$D_q$
satisfying \eqref{constraint8} is physically realizable if its
augmented  system \eqref{form4}-\eqref{form5} is physically
realizable, with explicit physical realizability conditions stated in Theorem \ref{theorem-1}. It is worthing noting that these physical realizability conditions are only suitable for an
augmented  system \eqref{form4}-\eqref{form5} with
$D_q=I$ or
$D_q=[\begin{array}{cc} I & 0 \end{array}]$
(no scattering processes involved). However, the matrix $D_q$ in the standard form system  \eqref{form4}-\eqref{form5} is allowed to be more general, namely, the one satisfying Eq. \eqref{constraint8}.  To deal with this, we need to extend the physical realizability condition of the system   \eqref{form4}-\eqref{form5}  by allowing a general matrix $D_q$ satisfying Eq. \eqref{constraint8}.  We first transform the
  augmented  system \eqref{form4}-\eqref{form5} into a familiar
form without scattering processes. Suppose that non-demolition condition
$[\tilde{x}(t), y_q (s) ^T]=0$,
$\forall$ $t\geq s \geq 0$ holds. So,  we apply  Eq. \eqref{new-conditionBCD} in Lemma
\ref{nondemolition0} to the quantum output $y_q$ in Eq. \eqref{form5} to get
$y_q=D_q\bar{y}_q$ with $\bar{y}_q$ defined as
$d\bar{y}_q=\bar{C}\tilde{x}(t)dt+dw(t)$, where  $\bar{C}=\Theta_{w}\tilde{B}^T\tilde{\Theta}$. Then, a
{\em reduced} system for the   augmented  system
\eqref{form4}-\eqref{form5} is defined as
\begin{eqnarray}
d\tilde{x}(t)&=&\tilde{A}\tilde{x}(t)dt+ \tilde{B}dw(t),\label{form6_a}\\
d\bar{y}_q&=&\bar{C}\tilde{x}(t)dt+dw(t)\label{form6_b}.
\end{eqnarray}It is straightforward to verify that
 the  reduced  system \eqref{form6_a}-\eqref{form6_b} is physically
realizable in the sense of Definition \ref{defn:mixed-qc-0} and satisfying the conditions of Theorem \ref{theorem-1}. The definition of physical realizability of an augmented  system of the system \eqref{standard-form1_a}-\eqref{standard-form1_b}  is as
follows:

\begin{definition}\label{pre-definition}
An  augmentation   \eqref{form4}-\eqref{form5} of the system
\eqref{standard-form1_a}-\eqref{standard-form1_b}  with a general matrix $D_q$ is said to be physically realizable if the
following statements hold:
\begin{enumerate}

\item The  reduced system \eqref{form6_a}-\eqref{form6_b} is physically realizable in the sense
of Definition \ref{defn:mixed-qc-0}.

\item For the augmented system \eqref{form4}-\eqref{form5}, non-demolition condition $[\tilde{x}(t), y_q(s)^T]=0, \  \forall\ t\geq s \geq 0$ holds.

\item $D_q$ is of the form $[\begin{array}{cc} I_{n_{y_q}} & 0 \end{array} ]\tilde{V}$ with $\tilde{V}$ a
symplectic matrix \cite{GJN2010} or unitary symplectic
\cite{NJD09} such that relation \eqref{constraint8} holds.
\end{enumerate}
\end{definition}

Next we will consider physical realizability conditions of the system
(\ref{form3}). Classical systems are always regarded as being
physically realizable since they can be approximately built via
digital and analog circuits. Thus, we just need to make sure that
output equation (\ref{form3}) is classical. Now, we can present a
formal definition of physical realizability of the system
\eqref{standard-form_a}-\eqref{standard-form_b}.

\begin{definition}\label{dfn:realize1}
A system of the  standard  form \eqref{standard-form_a}-\eqref{standard-form_b}   is said to
be physically realizable if the following statements hold:
\begin{enumerate}
\item There exists an   augmented  system \eqref{form4}-\eqref{form5} of
the system \eqref{standard-form1_a}-\eqref{standard-form1_b} with $D_q$ satisfying
\eqref{constraint8}, which is physically realizable in the sense of
Definition \ref{pre-definition}.

\item For the system \eqref{standard-form_a}-\eqref{standard-form_b}, non-demolition condition $[x(t), ~ y(s)^T]=0, \  \forall\ t\geq s \geq 0$ holds.

\item The output (\ref{form3}) and system variables $x_c$ both represent classical stochastic processes in the sense of  the following commutation relations $[x_c(t), ~x^T_c(s)]=0$, $[x_c(t),~
y^T_c(s)]=0$, and  $[y_c(t), ~y^T_c(s)]=0$ for all $t, s\geq0$.
\end{enumerate}
\end{definition}

The following theorem shows necessary and sufficient conditions for
physical realizability of system  \eqref{standard-form_a}-\eqref{standard-form_b}.

\begin{theorem}\label{Them-form}
A system of the form \eqref{standard-form_a}-\eqref{standard-form_b} is physically realizable,
if and only if matrices $A, B, C$, and $D$ satisfy the following
constraints: \begin{eqnarray} A\Theta_n+\Theta_n
A^{T}+B\Theta_{w}B^{T}=0\label{condition-1},\\
B\Theta_{w}D^T
=-\Theta_n
C^T\label{condition-BCD},\\
D\Theta_{w}D^T=\Theta_{y}.
\label{condition-D}
\end{eqnarray}
\end{theorem}

\quad \quad $\emph{Proof}$: ({\it Sufficiency}.) Let conditions
\eqref{condition-1}-\eqref{condition-D} hold. we proceed along the following  steps.

  $(i)$  Post-multiplying both sides of  \eqref{condition-BCD} by $\left[
                                                                                                    \begin{array}{c}
                                                                                                      I_{2n_{y_q}} \\
                                                                                                     0 \\
                                                                                                    \end{array}
                                                                                                  \right]$, we get
\begin{eqnarray} \label{proof:new-eq2}
B\Theta_{w}D_q^T
=-\Theta_n
C_q^T.
\end{eqnarray}It follows by
inspection that under conditions \eqref{condition-1} and \eqref{proof:new-eq2}, there exist
matrices $\tilde{A}, \tilde{B}, \tilde{C}$ and $\tilde{\Theta}$ satisfying the following conditions
\begin{eqnarray} \tilde{A}\tilde{\Theta}+\tilde{\Theta}
\tilde{A}^{T}+\tilde{B}\Theta_{w}\tilde{B}^{T}=0\label{proof:eq1},\\
\tilde{C}=D_q\Theta_{w}\tilde{B}^T\tilde{\Theta}\label{proof:eq2},
\end{eqnarray} where $A'$, $A''$, $B'$ are given by the following relations:
\begin{eqnarray}
B'\Theta_{w}D_q^T=[0\quad I]C_{q}^T, \\
\left[0 \quad I\right] A'^{T}-A'\left[
\begin{array}{c}
0 \\
I \\
\end{array}
\right]=B'\Theta_{w}B'^{T},\\
A''=\left(A'\Theta_n-[0\quad I]A^T+B'\Theta_{w}B^T\right)\left[
\begin{array}{c}
0 \\
I \\
\end{array}
\right].
\end{eqnarray}  From \eqref{form5} and  \eqref{proof:eq2}, we get
\begin{eqnarray}\label{proof:eq3}
\bar{C}=\Theta_{w}\tilde{B}^T\tilde{\Theta}.
\end{eqnarray}
Conditions \eqref{proof:eq1} and
\eqref{proof:eq3} imply the reduced  system \eqref{form6_a}-\eqref{form6_b}  satisfies the
physically realizability condition of Theorem \ref{theorem-1}. By  Lemma \ref{nondemolition0}, condition \eqref{proof:eq2} implies that $[\tilde{x}(t), y_q(s)^T]=0, \  \forall\ t\geq s \geq 0$ holds, which satisfies the second condition of  Definition \ref{pre-definition}.  Pre-multiplying and post-multiplying both sides of \eqref{condition-D} by $[\begin{array}{cc} I & 0 \end{array}]$ and $\left[
                           \begin{array}{c}
                             I \\
                             0
                           \end{array}
                         \right]$ respectively, we can obtain  \eqref{constraint8}. Thus,
the   augmented system (\ref{form4})-(\ref{form5}) is physically
realizable in the sense of Definition \ref{pre-definition}.

 $(ii)$ By Lemma \ref{nondemolition}, condition \eqref{condition-BCD} implies that $[x(t), y(s)^T]=0, \  \forall\ t\geq s \geq 0$ holds, which satisfies the second condition of  Definition \ref{dfn:realize1}.

 $(iii)$  Combining
conditions \eqref{constraint10},
\eqref{condition-BCD} and using the same approach as shown in the
proof of Lemma \ref{nondemolition0}, we get
$d_t[y_c(t),y_c(s)^T]=0,
d_t[y_c(s),y_c(t)^T]=0$ and
$d[y_c(t),y_c(t)^T]=0,$ for all
$t\geq s \geq 0$ (here the symbol $d_t$ denotes the forward differential with respect to $t$), which imply
that $[y_c(t),y_c(s)^T]=0$ holds for all $t,s \geq 0$ under the
fact that $[y_c(0), y_c(0)^T]=0$ given in Definition
\ref{dfn:realize1}. Applying a similar trick, we have
$[x_c(t),x_c(s)^T]=0, [
y_c(t),x_c(s)^T]=0 $ for all
$t,s\geq 0$. We infer that output (\ref{form3})
and $x_c$ are both classical in the sense of the third item
of Definition \ref{dfn:realize1}. Therefore, we conclude that the system
\eqref{standard-form_a}-\eqref{standard-form_b} is physically realizable in the sense of Definition \ref{dfn:realize1}, which shows that
\eqref{condition-1}-\eqref{condition-D} are sufficient for physical
realizability.

({\it Necessity}.)  Conversely, now suppose that a system of the form
\eqref{standard-form_a}-\eqref{standard-form_b} is physically realizable. It follows from
Theorem \ref{theorem-1} and the first item
of Definition \ref{dfn:realize1} that condition \eqref{proof:eq1} holds.
Then, reading off the first $n$ rows and columns of both sides of
\eqref{proof:eq1} gives us condition \eqref{condition-1}. By the
second item of Definition \ref{dfn:realize1}, we have condition
\eqref{condition-BCD} in the sense of Lemma \ref{nondemolition}. Since the system \eqref{standard-form_a}-\eqref{standard-form_b} is   a  standard  form, it follows from the third item of Definition
\ref{defn:standard-form} that condition \eqref{condition-D} holds.
Therefore, constraints \eqref{condition-1}-\eqref{condition-D} are
necessary for physical realizability. \hfill  $\blacksquare$

\subsection{Physical realizability for the general form}
In this subsection, we give an definition of the physical
realizability definition for the {\em general} form
\eqref{general-form_a}-\eqref{general-form_b}. A necessary and sufficient condition is also given.

\begin{definition}\label{defn:general}
A system of the {\em general} form \eqref{general-form_a}-\eqref{general-form_b} is said to
be physically realizable if its corresponding   standard  form
 \eqref{standard-form_a}-\eqref{standard-form_b} is physically realizable in the sense of
Definition \ref{dfn:realize1}.
\end{definition}

\begin{theorem}\label{Th:general-form}
A system of the {\em general} form \eqref{general-form_a}-\eqref{general-form_b} is
physically realizable, if and only if the following constraints are
satisfied: \vspace{-1em} \begin{eqnarray}
\mathbf{A}\hat{\Theta}_n+\hat{\Theta}_n
\mathbf{A}^{T}+\mathbf{B}\Theta_v\mathbf{B}^{T}=0 \label{g-condition1},\\
\mathbf{B}\Theta_v\mathbf{D}^T
=-\hat{\Theta}_n
\mathbf{C}^T\label{g-conditionBCD},\\
\mathbf{D}\Theta_v\mathbf{D}^T=\Theta_\mathbf{y}\label{g-conditionD}.
\end{eqnarray}
\end{theorem}
\quad \quad $\emph{Proof}$:
Suppose that equations (\ref{g-condition1})-(\ref{g-conditionD})
hold. It follows from Theorem \ref{two-relation} that the general
system \eqref{general-form_a}-\eqref{general-form_b} can be transformed to its corresponding
 standard  system \eqref{standard-form_a}-\eqref{standard-form_b}. Using relations
\eqref{relations} and equations
(\ref{g-condition1})-(\ref{g-conditionD}), we get constraints
\eqref{condition-1}-(\ref{condition-D}). The corresponding  standard  system
 \eqref{standard-form_a}-\eqref{standard-form_b} is physically realizable in the sense of Theorem
\ref{Them-form}. Therefore, we
conclude that (\ref{g-condition1})-(\ref{g-conditionD}) are
sufficient for physical realizability.

Conversely, suppose that a system of the general form
\eqref{general-form_a}-\eqref{general-form_b} is physically realizable. It follows from
Definition \ref{defn:general} and Theorem \ref{Them-form} that
constraints \eqref{condition-1}-(\ref{condition-D}) hold. Conditions
(\ref{g-condition1})-(\ref{g-conditionD}) can be obtained from
constraints \eqref{condition-1}-(\ref{condition-D}) by direct
substitution using relations \eqref{relations}. Thus, constraints
(\ref{g-condition1})-(\ref{g-conditionD}) are necessary for
realizability.
\hfill  $\blacksquare$

\section{Systematic synthesis of mixed quantum-classical linear stochastic systems}\label{sec:them}

By Theorem \ref{two-relation} and Definition \ref{defn:general}, we know that
a system of the general form \eqref{general-form_a}-\eqref{general-form_b} can be physically
realized, if its corresponding standard  form
 \eqref{standard-form_a}-\eqref{standard-form_b} is physically realizable. Therefore, our
purpose in this section is to develop a network synthesis theory
only for a mixed quantum-classical system of the   standard  form
 \eqref{standard-form_a}-\eqref{standard-form_b} that generalizes the results in \cite{Nurdin2011}.

\begin{lemma}\label{conditions}
The mixed quantum-classical linear stochastic system
\eqref{standard-form_a}-\eqref{standard-form_b} is physically realizable if and only if
conditions \eqref{constraint8}-\eqref{constraint10} and the constraints below are all satisfied
\begin{eqnarray}
A_{qq}\Theta_{n_q}+\Theta_{n_q}A_{qq}^T+B_q\Theta_{w}B_q^T=0\label{constraint1},\\
A_{cq}\Theta_{n_q}+B_c\Theta_{w}B_q^T=0\label{constraint2},\\
B_c\Theta_{w}B_c^T=0\label{constraint3},\\
B_c\Theta_{w}D_q^T=0\label{constraint4},\\
B_q\Theta_{w}D_q^T=-\Theta_{n_q}C_{qq}^T\label{constraint5},\\
B_c\Theta_{w}D_c^T=0\label{constraint6},\\
B_q\Theta_{w}D_c^T=-\Theta_{n_q}C_{cq}^T\label{constraint7}.
\end{eqnarray}
\end{lemma}

\quad \quad $\emph{Proof}$:
By Theorem \ref{Them-form}, it is easily checked that  conditions \eqref{constraint8}-\eqref{constraint10} are equivalent to \eqref{condition-D} while \eqref{constraint1}-\eqref{constraint7}  are equivalent to  \eqref{condition-1}-\eqref{condition-BCD}.
\hfill $\blacksquare$

\begin{lemma}\label{lemma-D}
If a matrix $D_q$ satisfies the condition
\begin{eqnarray}
D_{q}\Theta_{w}
D_{q}^T&=&\Theta_{y_q}\label{p1},
\end{eqnarray} then there exists a matrix
$D_{q}'$ such that
\begin{equation}\label{eq:jun16_Dqq'}
\left[
\begin{array}{c}
D_{q} \\
D_{q}' \\
\end{array}
\right]\Theta_{w}\left[
\begin{array}{c}
D_{q} \\
D_{q}' \\
\end{array}
\right]^T=\Theta_{w}.
\end{equation}
\end{lemma}

\quad \quad $\emph{Proof}$:
The matrix $D_q$ can be written in the form of
{\setlength\arraycolsep{3pt}\begin{eqnarray}\label{G1}
D_{q}=\left[
\begin{array}{cc}
I & 0_{2n_{y_q}\times (2m-2n_{y_q})} \\
\end{array}
\right]\left[
\begin{array}{c}
D_{q} \\
D_{q}' \\
\end{array}
\right],
\end{eqnarray}}where $D_{q}'$ is a
$(2m-2n_{y_q})\times 2m$
matrix to be constructed. Let the rows of $D_{q}$ be denoted by $d_1,d_2, \cdots,
d_{2n_{y_q}}$.
Let $P(a| b_1,b_2,\cdots, b_k)$ denote the
orthogonal projection of the row vector $a$ onto
the subspace spanned by the row vectors $b_1,b_2,\cdots,
b_k$. Now, we build a $(2m-2n_{y_q})\times 2m$ matrix $D_{q}'$, following analogously the construction of the matrix $V$ defined in
\cite[Lemma 6]{Nurdin2011}. First, choose a row vector $v^{(1)}_{ 1}\in
\mathbb{R}^{2m}$ linearly independent of $d_1,d_2,\cdots, d_{2n_{y_q}}$, and set $v^{(2)}_{ 1}=v^{(1)}_{ 1}-P(v^{(1)}_{ 1}|d_1,d_2,\cdots, d_{2n_{y_q}})$ and $v_{ 1}=v^{(2)}_{ 1}\Theta_{w}$. Next, choose a row vector $v^{(1)}_{ 2}\in
\mathbb{R}^{2m}$ linearly independent of $d_1,d_2,\cdots, d_{2n_{y_q}}$ and set $v^{(2)}_{ 2}=v^{(1)}_{ 2}-P(v^{(1)}_{ 2}|d_1,d_2,\cdots, \\ d_{2n_{y_q}}, v_1)$ and $v_{ 2}=v^{(2)}_{ 2}\Theta_{w}$. Repeat this procedure analogously for $k= 3, \cdots, m-n_{y_q}$ to obtain vectors $v_{ k}=v^{(k)}_{ k}\Theta_{w}$ with $v^{(k)}_{ k}=v^{(k-1)}_{ k}-P(v^{(k-1)}_{ k}|d_1,d_2,\cdots, d_{2n_{y_q}}, v_1, \cdots, v_{k-1})$.
Then, we choose a row vector $w^{(1)}_{1}\in
\mathbb{R}^{2m}$ that is linearly independent of $d_1,d_2,\cdots, d_{2n_{y_q}}$ and $v_2, v_3, \cdots,
v_{m-n_{y_q}}$ such that $(w^{(1)}_{1}-P(w^{(1)}_{1}|d_1,d_2,\cdots, d_{2n_{y_q}}$, $v_2, v_3, \cdots,
v_{m-n_{y_q}}))v_1^T\neq0$. Set
$w_1^{(2)}=w_1^{(1)}-P(w^{(1)}_{1}|d_1,d_2,\cdots, d_{2n_{y_q}}$, $v_2, v_3, \cdots,
v_{m-n_{y_q}})$ and $w_1=w_1^{(2)}\Theta_{w}/(v_1w_{1}^{(2)T})$.
Next, we choose
$w^{(1)}_{2}\in\mathbb{R}^{2m}$
that is linearly independent of
$d_1,d_2,\cdots, d_{2n_{y_q}}$ and $v_1,
w_1, v_3, v_4,\cdots, v_{m-n_{y_q}}$ such that
$(w^{(1)}_{1}\!-\!P(w^{(1)}_{1}|d_1,d_2,\cdots,
d_{2n_{y_q}}$, $v_1, w_1, v_3,
v_4\cdots,
v_{m-n_{y_q}}))v_2^T\neq0$. Set
$w_2^{(2)}=w_2^{(1)}\!-\!P(w^{(1)}_{2}|d_1,d_2$, $\cdots, d_{2n_{y_q}}, v_1, w_1, v_3, v_4,\cdots,
v_{m-n_{y_q}})$ and
$w_2=w_2^{(2)}\Theta_{w}/(v_2w_{2}^{(2)T})$. Repeat
the procedure in an analogous manner to construct
$w_3, w_4, \cdots$, $w_{m-n_{y_q}}$. Then the matrix $D_{q}'$ is defined as
\begin{eqnarray}\label{V'}
D_{q}'
=[v_1^T, w_1^T, v_2^T, w_2^T, \cdots, v_{m-n_{y_q}}^T,
w_{m-n_{y_q}}^T]^T \in \mathbb{R}^{2(m-n_{y_q})\times 2m}.
\end{eqnarray}
By the construction above it is clear that Eq. \eqref{eq:jun16_Dqq'} holds.
\hfill $\blacksquare$

\begin{remark} \label{rem:tildeV}
According to Eq. \eqref{eq:jun16_Dqq'}, the matrix $D_q$ can be embedded into a symplectic
matrix
\begin{equation} \label{tildeV}
\tilde{V}=\left[
\begin{array}{c}
D_{q}\\
D_{q}'
\end{array}
\right] \in \Bbb{R}^{2m\times 2m}
\end{equation}
 which satisfies  $\tilde{V}\Theta_{w}\tilde{V}^T=\Theta_{w}$.
\end{remark}

Suppose that the system \eqref{standard-form_a}-\eqref{standard-form_b}, or equivalently system \eqref{form1}-\eqref{form3}, is physically
realizable. We are now in a position to explain how to realize the system
 \eqref{standard-form_a}-\eqref{standard-form_b} as an interconnection of  a classical system $G_1$ described by \eqref{classical-eq1}-\eqref{classical-eq4} and a quantum system $G_2$ described by \eqref{quantum-eq1}-\eqref{quantum-eq3}. To do this, we have to determines the system matrices for $G_1$ and $G_2$. Notice that $A_{qq}, B_q, C_{qq},D_q$ are already given in Eq. \eqref{ABCD} for system \eqref{form1}-\eqref{form3}, all the undetermined matrices are those with superscript $'$. In what follows we show how they can all be determined under the assumption of the physical realizability of the system  \eqref{standard-form_a}-\eqref{standard-form_b}.

First of all, in analogy to the partitioning of $w(t)$ in Subsection \ref{s-form}, we partition $ w'(t)=\left[\begin{array}{c} w'_1(t) \\ w'_2(t) \\ \end{array}\right]$,  where $w'_k(t)$ is of the same dimension as $w_k(t)$, ($k=1,2$).

Secondly, by Lemma \ref{lemma-D}, the matrix $D'_q$ in Eq. \eqref{quantum-eq3} can be constructed.

Thirdly, the matrix $C'_{qq}$ in Eq. \eqref{quantum-eq3} can be constructed by means of
 \begin{equation} \label{eq:jun16_Cqq'}
 C'_{qq}=D'_q\Theta_{w}B_{q}^T\Theta_{n_q}.
\end{equation}

Finally, the remaining undefined system matrices, input and output signals appearing in   \eqref{classical-eq1}-\eqref{quantum-eq3} can be found in the following theorem, which also   presents  a feedback architecture for the realization of the system  \eqref{standard-form_a}-\eqref{standard-form_b}.

\begin{figure}[htbp]
\vspace{-9em}\centering
\includegraphics[height=155mm,width=120mm]{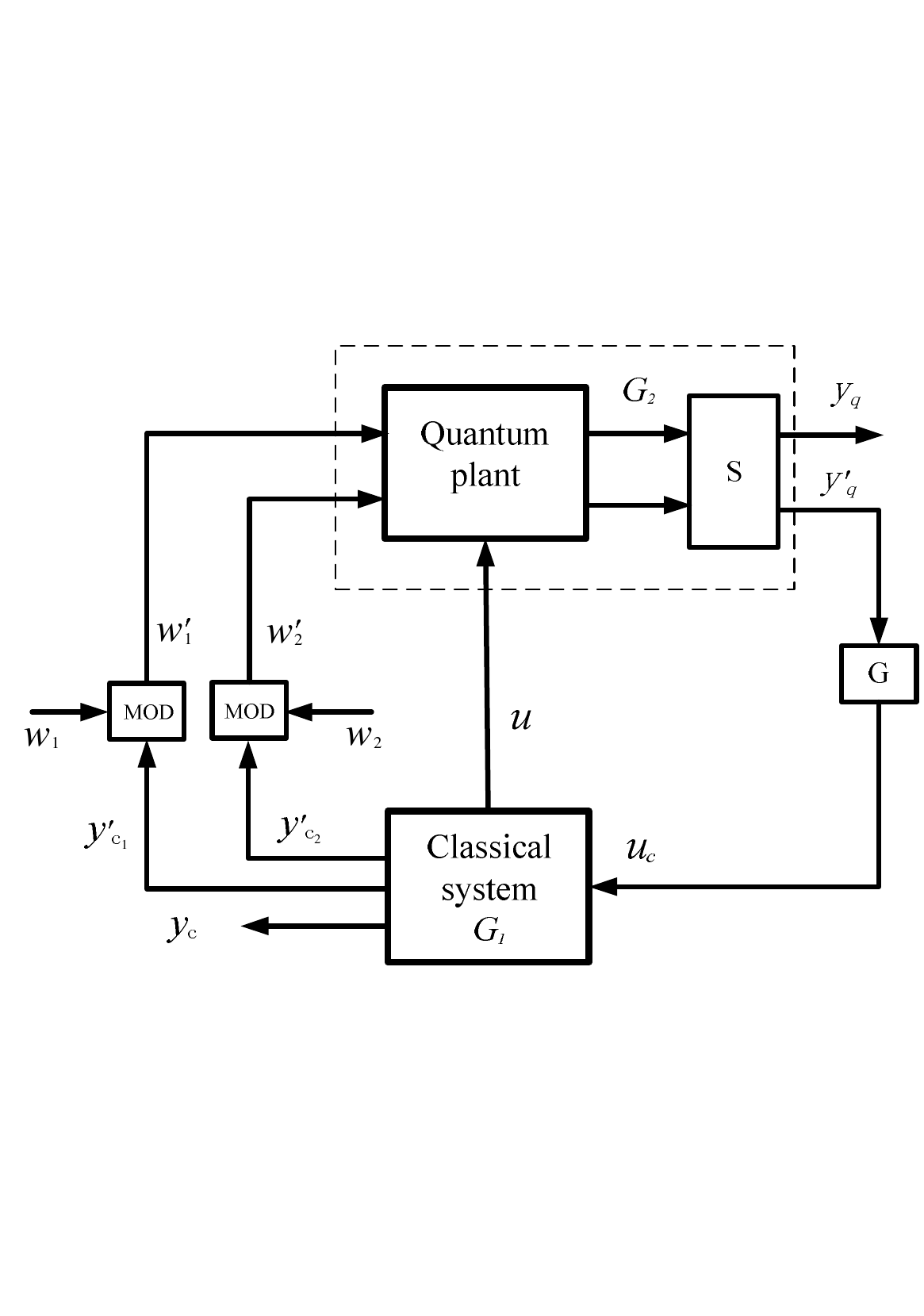}
\vspace{-10em}\caption{Feedback interconnection of  a classical system $G_1$ and a quantum system
$G_2$. \label{fig:q-c-loop}}
\end{figure}

\begin{theorem} \label{thm:main}
Assume that the system \eqref{standard-form_a}-\eqref{standard-form_b}, or equivalently system \eqref{form1}-\eqref{form3} with system matrices given in Eq. \eqref{ABCD},  is physically realizable and all its system matrices are already known.  Then there  exist matrices
$C'_{c}\equiv \left[\begin{array}{c}
C_{c_1}' \\
C_{c_2}'
\end{array}
\right]$, $G$, $B_c'$, and
$D_c'$, such that
\begin{eqnarray}
D_{q}C'_{c}&=&C_{qc}\label{main:eq3},\\
B_{c}'GC'_{qq}&=&A_{cq}\label{main:eq8},\\
B_{c}'GD'_{q}&=&B_{c}\label{main:eq9},\\
D_{c}'GC'_{qq}&=&C_{cq}\label{main:eq10},\\
D_{c}'GD'_{q}&=&D_{c}\label{main:eq11}.
\end{eqnarray}
Moreover, a feedback network realization of the system \eqref{standard-form_a}-\eqref{standard-form_b} shown in Figure \ref{fig:q-c-loop}\footnote[1]{The two sets of modulators
(MODs) presented in Figure \ref{fig:q-c-loop} displace the vectors of vacuum quantum fields $w_1$ and $w_2$
to produce the quantum signals $w'_1(t)$ and $w'_2(t)$ by the
classical vector signals
$y'_{c_1}(t)$ and
$y'_{c_2}(t)$, respectively.}, with the identification
\begin{eqnarray}
E &=&A_{qc}-B_{q}C'_{c},
\label{jun25_E}
\\
A_{cc}' &=&A_{cc}-B_c'GD'_qC'_{c},
\label{jun25_Acc'}
\\
C_{cc}'
 &=& C_{cc}-D_c'GD'_qC'_{c},
 \label{jun25_Ccc}
\\
u(t) &=& x_c(t),
\label{jun25_u}
\\
u_c(t) &=& G y'_q(t),
\label{jun25_uc}
\\
w'_1(t) &=&y_{c_1}'(0)+\int^t_0y_{c_1}'(s)ds+w_1(t),
\label{jun25_w1'}
\\
w'_2(t) &=&y_{c_2}'(0)+\int^t_0y_{c_2}'(s)ds+w_2(t),
\label{jun25_w2'}
\end{eqnarray}
 is a
physical realization of the system \eqref{standard-form_a}-\eqref{standard-form_b} consisting
of  a classical system $G_1$
described by \eqref{classical-eq1}-\eqref{classical-eq4} and a quantum system $G_2$ described by
\eqref{quantum-eq1}-\eqref{quantum-eq3}. The network $G$  in Figure \ref{fig:q-c-loop}, which corresponds to measurement processes, can realize the matrix $G=KV$ (to be given in Eq. \eqref{GBD} below) to produce classical signals
$u_c=Gy_q'(t)$ satisfying
$[u_c(t), ~ u_c(s)^T]=0, \forall
t,s\geq 0$; the network S realizes the symplectic transformation
$\tilde{V}$ in Eq. \eqref{tildeV}.
\end{theorem}

\quad \quad $\emph{Proof}$: The proof consists of the following six steps.

{\it Step 1.} Construct the matrix $C'_{c}$ satisfying Eq. \eqref{main:eq3}.  It follows from Eq.
\eqref{constraint8} with an invertible $\Theta_{y_q}$ that the
matrix $D_{q}$ has full row rank and thus $\mathrm{rank}(\!D_{q}\!)\!=\!\mathrm{rank}\left(\! [
\begin{array}{cc}
D_{q}\quad C_{qc} \\
\end{array}
\!]\right)$.
Consequently, the solution of Eq. \eqref{main:eq3} can be given as
$C'_{c}=D_q^T(D_qD_q^T)^{-1}C_{qc}+N(D_q)$,
where $N(D_q)$ denotes a matrix of the same dimension as $C'_{c}$ whose columns are in the kernel space of $D_q$.

{\it Step 2.} Let
\begin{equation} \label{eq:jun16_BDc}
\bar{B}_c=B_c'G, ~~
\bar{D}_c=D_c'G.
\end{equation}
 Then Eqs. \eqref{main:eq9} and \eqref{main:eq11} can be re-written as
\begin{equation} \label{eq:jun16_barbc}
\left[
\begin{array}{c}
\bar{B}_c \\
\bar{D}_c\\
\end{array}
\right]D'_q=\left[
\begin{array}{c}
B_{c} \\
D_{c} \\
\end{array}
\right].
\end{equation}
We show that Eq. \eqref{eq:jun16_barbc} has a solution $\left[
\begin{array}{c}
\bar{B}_c \\
\bar{D}_c \\
\end{array}
\right]$.
Combining Eqs. \eqref{constraint9}, \eqref{constraint4} and  \eqref{eq:jun16_Dqq'}
gives {\setlength\arraycolsep{2pt}\begin{eqnarray} \left[
\begin{array}{c}
D_{q} \\
D'_q \\
\end{array}
\right]\Theta_{w}(D'_q)^T&=&\left[
\begin{array}{c}
0_{2n_{y_q}\times (2m-2n_{y_q})} \\
\Theta_{y_q'}\\
\end{array}
\right]\label{p-main1},
\\
\left[
\begin{array}{c}
D_{q} \\
D'_q \\
\end{array}
\right]\Theta_{w}\left[
\begin{array}{cc}
B_{c}^T \quad D_{c}^T \\
\end{array}
\right]&=&\left[
\begin{array}{c}
0_{2n_{y_q}\times (n_c+n_{y_c})} \\
D'_q\Theta_{w}\left[
\begin{array}{cc}
B_{c}^T &D_{c}^T \\
\end{array}
\right] \\
\end{array}
\right]\label{p-main2},
\end{eqnarray}}where
$\Theta_{y_q'}=\mathrm{diag}_{(m-n_{y_q})}(J)$.
From equations \eqref{p-main1} and \eqref{p-main2}, we can infer
that
{\setlength\arraycolsep{1pt}$\mathrm{rank}\left((D'_q)^T\right)=\mathrm{rank}(\Theta_{y_q'})$
, $\mathrm{rank}\left([
\begin{array}{ccc}
B_{c}^T \quad D_{c}^T \\
\end{array}
]\right)=\mathrm{rank}\left(D'_q\Theta_{w}[
\begin{array}{cc}
B_{c}^T \quad D_{c}^T \\
\end{array}]\right)$}. Given that $\Theta_{y_q'}$ has full row rank, we can conclude that {\setlength\arraycolsep{1pt}$\mathrm{rank}\left(\Theta_{y_q'}\right)=\mathrm{rank}\left(\left[\Theta_{y_q'} \quad D'_q\Theta_{w}[
\begin{array}{cc}
B_{c}^T \quad D_{c}^T \\
\end{array}]\right]\right)$}, which implies that {\setlength\arraycolsep{1pt}$\mathrm{rank}\left((D'_q)^T\right)=\mathrm{rank}\left(\left[
\begin{array}{ccc}
(D'_q)^T \quad B_{c}^T \quad D_{c}^T \\
\end{array}
\right]
\right)$}. So, there exist
$\bar{B}_c$ and
$\bar{D}_c$ satisfying Eq. \eqref{eq:jun16_barbc}.

{\it Step 3.} We construct matrices
$C'_{c}$, $G$, and $B_c'$.  We get from equations (\ref{constraint10}), (\ref{constraint3}),
(\ref{constraint6}), (\ref{main:eq9}), (\ref{main:eq11}), and (\ref{p-main1}) that
\begin{eqnarray}\label{main:eq12}
\left[
\begin{array}{c}
\bar{B}_c \\
\bar{D}_c\\
\end{array}
\right] \Theta_{y_q'}\left[
\begin{array}{c}
\bar{B}_c \\
\bar{D}_c\\
\end{array}
\right]^T=0.
\end{eqnarray}From equation (\ref{main:eq12}), we know that the matrix
$\left[
\begin{array}{c}
\bar{B}_c \\
\bar{D}_c\\
\end{array}
\right]$ with
$\mathrm{rank}\left(\left[
\begin{array}{c}
\bar{B}_c \\
\bar{D}_c\\
\end{array}
\right]\right)=r$ can be decomposed as
\[\left[
\begin{array}{c}
\bar{B}_c \\
\bar{D}_c\\
\end{array}
\right]=PZKV=\left[
\begin{array}{c}
P_1Z \\
P_2Z \\
\end{array}
\right]KV,
\]
 where $P=\left[
\begin{array}{c}
P_1 \\
P_2 \\
\end{array}
\right]$ is a permutation matrix; $Z$ is a matrix of the form
$Z=\left[
\begin{array}{c}
I_r \\
X\\
\end{array}
\right]
$ if $r<n_c+n_{y_c}$, where $X$ is some
$(n_c+n_{y_c}-r)\times r$
matrix, $Z=I_{(n_c+n_{y_c})}$ if
$r=n_c+n_{y_c}$,
\begin{equation} \label{eq:jun18_K}
K= \left[
\begin{array}{c}
k_1 \\
k_2\\
\vdots\\
k_{r}
\end{array}
\right]
=
\left[
\begin{array}{cc}
I_r & 0_{r \times (n_{y_q'}-r)}
\end{array}
\right] \in \mathbb{R}^{r\times n_{y_q'}},
\end{equation}
 and
$V$ is a symplectic matrix
 (see \cite[Lemma 6]{Nurdin2011} for
details). Being symplectic,   the matrix $V$ can be realized as a suitable static quantum optical network, \cite{LN2004}. We
define
\begin{equation} \label{GBD}
 G=KV,\quad B_c'=P_1Z, \quad D_c'=P_2Z.
\end{equation}


{\it Step 4.} From
Eqs. \eqref{constraint2}, \eqref{constraint7}, and $C'_{qq}$ defined in Eq. \eqref{eq:jun16_Cqq'}, we conclude
that Eq. \eqref{main:eq9} implies Eq. \eqref{main:eq8}, and Eq.
\ref{main:eq11} implies  Eq. \eqref{main:eq10}, respectively.

{\it Step 5.} It is straightforward to verify from Eqs.
\eqref{main:eq3}-\eqref{jun25_w2'} that interconnecting the classical
system $G_1$ and the quantum system $G_2$
gives the  standard  form \eqref{standard-form_a}-\eqref{standard-form_b}, or equivalently described
by \eqref{form1}-\eqref{form3}. Now let us check that the system
$G_2$ is a physically realizable fully quantum system. It follows
from conditions \eqref{constraint8} and \eqref{constraint1} that the
system $G_2$ satisfies constraints \eqref{condition-1} and
\eqref{condition-D} in the sense of Theorem \ref{Them-form} with
matrices
$A$, $B$, $D$, $\Theta_{n}$ and
$\mathrm{diag}(\Theta_{y_q},
0_{n_{y_c}\times n_{y_c}})$ replaced by corresponding matrices
$A_{qq}$, $B_q$, $D_q$,
$\Theta_{n_q}$ and
$\Theta_{y_q}$, respectively.
The system $G_2$ also satisfies constraint \eqref{condition-BCD}
with its matrices replaced by corresponding matrices in equations
\eqref{classical-eq1}-\eqref{classical-eq4} with the proof as
follows: \begin{eqnarray*} -\Theta_{n_q}\left(\left[
\begin{array}{c}
D_{q}\\
D'_q \\
\end{array}
\right]\Theta_{w}B_{q}^T\Theta_{n_q}\right)^T&=&-\Theta_{n_q}\Theta_{n_q}^TB_{q}\Theta_{w}^T\left[
\begin{array}{c}
D_{q}\\
D'_q \\
\end{array}
\right]^T
=B_{q}\Theta_{w}\left[
\begin{array}{c}
D_{q}\\
D'_q \\
\end{array}
\right]^T.
\end{eqnarray*}So, the system $G_2$ is a physically realizable
quantum system, where $y_q'$ is the input to the network G.

{\it Step 6.} By Eqs. \eqref{eq:jun18_K} and \eqref{GBD},  Applying
$K$ to
$Vy'_q(t)$ is to measure the
first $r$ amplitude quadrature components of
$Vy'_q(t)$ to obtain the
measurement
result $u_c(t)=KVy'_q(t) = Gy'_q(t)$. So,
$G$ represents measurement
processes, \cite{Nurdin2011},  \cite{NY17}. Then we can show that
\begin{eqnarray*}
[u_c(t), ~u_c(s)^T]=G[y_q'(t),~y_q'(s)^T]G^T=2i\delta_{ts}G\Theta_{y_q'}G^T=2i\delta_{ts}\times0=0, \forall
t,s\geq 0,
\end{eqnarray*} which implies that $u_c$ is classical. Thus $G_1$ described by  \eqref{classical-eq1}-\eqref{classical-eq4} is a classical system, where the
classical vector signals
$y'_{c_1}(t)$ and
$y'_{c_2}(t)$ are used to modulate $w_1(t)$ and $w_2(t)$ to produce the quantum signals $w'_1(t)$ and $w'_2(t)$ which are then injected into $G_2$.
\hfill $\blacksquare$

\begin{remark} \label{rem:jun25_ww}
By Eqs. \eqref{jun25_w1'}-\eqref{jun25_w2'}, we have
\[
 [dw'_1(t), ~ dw'_2(t)^T] = [dw_1(t),~   dw_2(t)^T].
\]
\end{remark}

\section{Application}
\label{sec:application}
When a system is described by a certain mathematical model,  it is often important to perform some form of analysis on it. Our paper provides a mathematical means to convert a general representation of a  mixed system to a standard form in which the system can be decomposed into two subsystems  which make clear the quantum and classical components of the system. The main results of this paper may thus have a practical application in  the analysis of measurement-based feedback control of quantum systems  described by LSDEs, where the plant is a quantum system while the
controller is a classical system \cite{WM10}, \cite{Gough12}, \cite{Yama14}, \cite{WSPSGK15}, \cite{NY17}. This decomposition results in the mixed system with a more illuminating structure, making it easier to draw conclusions on the system's quantum and classical subspaces. Then the quantum  subsystem can be  synthesized  by quantum optical devices like beam splitters, phase shifters, optical cavities, squeezers, etc, and the classical subsystem can be built by  standard analog or digital electronics;
see \cite{NJP09}, \cite{WNZJ13}, \cite{NY17}. Now an example is given to  illustrate  our main results.

$\mathbf{Example \  2}$: Consider a mixed quantum-classical system of the
standard  form with
$A, B,
C, D$ satisfying the physical realizability conditions
\eqref{condition-1}-\eqref{condition-D},
\begin{eqnarray*}
&&A=\left[
\begin{array}{ccc}
-9& -3 & -1\\
1 & -7 & -3\\
-0.72 & -0.6 & -12\\
\end{array}
\right]
,\quad B=\left[
\begin{array}{cccccc}
1 & 2 & -7 & 0 &-3 & 5\\
2 & 5 & 1 & -3 & 6 & -8\\
0 & 0.12 & 0 & 0 & 0 & -0.16\\
\end{array}
\right],\\
&&C=\left[ \begin{array}{ccc}
    38   &      46&  -42\\
         0.31     &   0.4  &  0.35\\
         4.2       &  -6 &   5\\
\end{array}
\right],\quad D=\left[
\begin{array}{cccccc}
    8&       0&   10&       0  &  6&        0\\
         0  &  0.04&       0  &  0.05&       0  &  0.03\\
    0&       0.8   &      0    &     -1 & 0&   0.6\\
\end{array}
\right].
\end{eqnarray*}
Following the construction in the proof of Theorem \ref{thm:main},
we have the classical system $G_1$ described by
\begin{eqnarray*}
dx_c(t)&=&-12x_c(t)+[ 3.6836  \quad -0.4345]du_c(t)\label{classical-eq1-1},\\
dy_c(t)&=& 12x_c(t)dt+
[ -0.2065   \quad 1.2388]du_c(t)\label{classical-eq2-1}, \\
y_{c_1}'(t)&=&0,\\ y_{c_2}'(t)&=&\left[
\begin{array}{c}
-4.2 \\
7\\
 0\\
 0
\end{array}
\right]x_c(t),
\end{eqnarray*}
 the quantum system $G_2$ given by
\begin{eqnarray*}\label{quantumeq1}
\hspace{-4em}dx_q(t)&=&\left[
\begin{array}{cc}
-9& -3 \\
1& -7 \\
\end{array}
\right]
x_q(t)dt\!+\!\left[
\begin{array}{cc|cccc}
1 & 2 & -7 & 0 & -3 & 5\\
2& 5 & 1 & -3 & 6 & -8\\
\end{array}
\right]\!\!\left[
\begin{array}{c}
dw_1(t) \\
dw_2(t) \\
\end{array}
\right]\!+\!\left[
\begin{array}{c}
-30.4\\
22.2\\
\end{array}
\right]du(t),\\
dy_q(t)&=&\left[
\begin{array}{cc}
38 & 46\\
0.31 & 0.4\\
\end{array}
\right]x_q(t)dt+\left[
\begin{array}{cc|cccc}
    8&       0&   10&       0  &  6&        0\\
         0  &  0.04&       0  &  0.05&       0  &  0.03\\
\end{array}
\right]\left[
\begin{array}{c}
dw_1(t) \\
dw_2(t) \\
\end{array}
\right],\\
dy_q'(t)&=&\left[
             \begin{array}{c}
               dy_{q_1}'(t) \\
               dy_{q_2}'(t) \\
             \end{array}
           \right]
=\left[
\begin{array}{cc}
-1.1 & 2.3\\
4.2& -6\\
-47& -14\\
-0.72 &-0.6\\
\end{array}
\right]
x_q(t)dt\!+\!\left[\!\!\!
\begin{array}{cc|cccc}
    0.4   &  0  & -0.5      &   0&    0.3&        0\\
         0    &0.8      &  0  & -1&         0&    0.6\\
3 & 0 &  0     &    0 & -4 & 0\\
0  &  0.12&       0    &     0    &     0 &  -0.16\\
\end{array}
\!\!\!\right]\left[
\begin{array}{c}
dw_1(t) \\
dw_2(t) \\
\end{array}
\!\!\right],
\end{eqnarray*}
and  the matrix $G$ is given by
\[
G\!=\!\left[\!\begin{array}{cccc}
0& 0.0971 & 0 & 0.2769\\
0&  0.8235 & 0 & 0.0462\\
\end{array}
\!\right].
\]

It can be easily checked that the  closed-loop system described by  \eqref{standard-form_a}-\eqref{standard-form_b} with the above matrices $A$, $B$, $C$, $D$ is obtained by making the identification
\begin{eqnarray*}u(t)&=&x_c(t),\\
du_c(t)
&=&\left[
\begin{array}{cc}
 0.2086 &  -0.7489\\
    3.4253   &-4.9684\\
\end{array}
\right]x_q(t)dt+\!\!\left[\!
\begin{array}{cc|cccc}
0& 0.1109   & 0  & -0.0971   & 0  &  0.014\\
0 &   0.6643  &       0 &  -0.8235   &      0 &   0.4867\\
\end{array}\!\right]\!\!
\!\left[\begin{array}{c}
dw_1(t) \\
dw_2(t) \\
\end{array}
\right],\\
dw'_1(t)&=&dw_1(t),\\
dw'_2(t)&=&\left[
\begin{array}{c}
-4.2\\
7\\
0\\
0
\end{array}
\right] x_c(t)dt\!+\! dw_2(t).
\end{eqnarray*}

\begin{figure}[htbp]\vspace{-24em}
\centering
  \includegraphics[height=240mm,width=185mm]{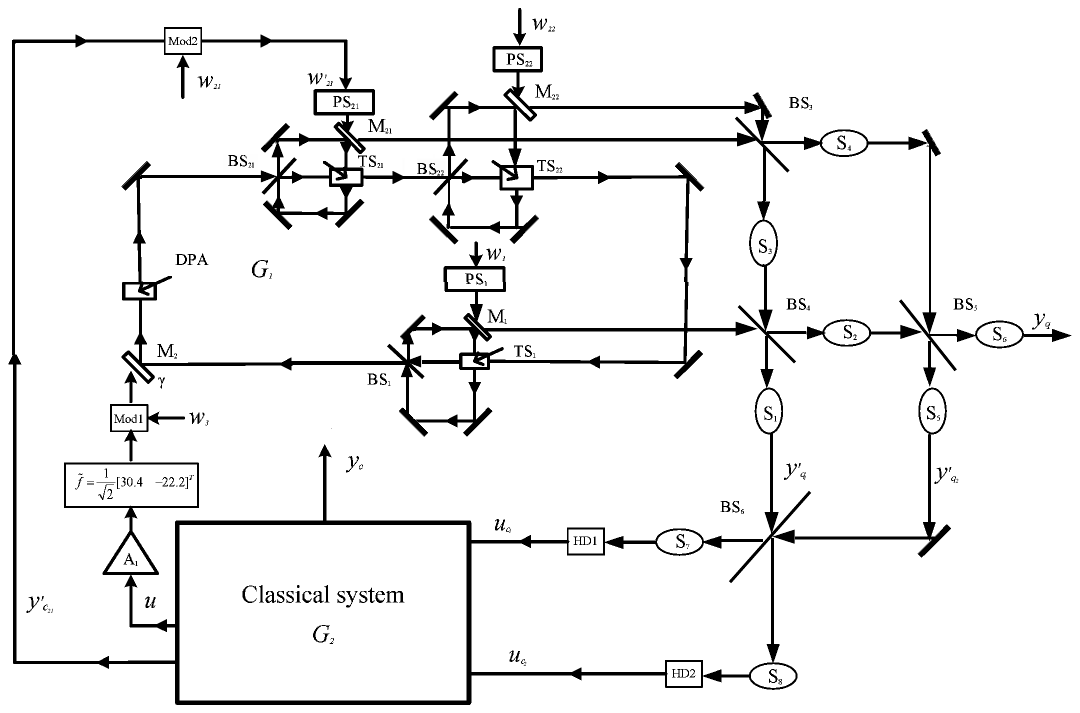}
 \vspace{-13em}\caption{A realzation of the mixed quantum-classical system in {\bf Example 2}. Black rectangles denote fully reflecting mirrors. $\mathrm{M_1}, \mathrm{M_{21}}, \mathrm{M_{22}}$ and $\mathrm{M_{3}}$ represent transmitting mirrors with coupling constants $\kappa_1, \kappa_{21}, \kappa_{22}$ and $\gamma$, respectively $(\gamma\ll1, \gamma\ll \kappa_1, \kappa_{21}, \kappa_{22})$; $\mathrm{BS_1}, \mathrm{BS_{21}}, \mathrm{BS_{22}}, \mathrm{BS_3}, \mathrm{BS_4}, \mathrm{BS_5}$ and $\mathrm{BS_6}$ represent beam splitters; $\mathrm{TS_1}, \mathrm{TS_{21}}$ and $\mathrm{TS_{22}}$ represent two-mode squeezers; $\mathrm{PS_1}$, $\mathrm{PS_{21}}$, $\mathrm{PS_{22}}$ represent phase shifters; $\mathrm{S_i}$ $(i=1,2\cdots, 8)$ represents a squeezer; DPA is short for degenerate parametric
amplifier; $\mathrm{Modi}$ $(i=1, 2, 3, 4)$ represents a modulator; $\mathrm{HDi}$ $(i=1,2)$ represents a homodyne detector; $\mathrm{A_1}$ is a amplifier with gain $1/\sqrt{\gamma}$. $\tilde{f}$ can be realized using a computer. $w_1, w_{21}, w_{22}, w_{3}$ are vacuum noises and the contribution of $w_3$ to quantum system noise is negligible compared to that of other vacuum noises. $G_2$ can be realized by  electrical and electronic devices, see \cite{AV1973}.
\label{fig:real5}}
\end{figure}

  The realization of this mixed system is
shown in Figure \ref{fig:real5}. The details of the construction and the individual components involved can be
found in \cite{JNP08}, \cite{NJD09}, \cite{WNZJ13},  \cite{NY17} and the references therein.

\section{Conclusion}
\label{sec:conclusion} In this paper, we explicitly detail how to obtain  mathematical representations for  a class of mixed quantum-classical linear stochastic systems; two forms (a   standard form and
a general form) are presented for the physical realization of
such mixed systems. We have also established
the relation between these two forms. Three physical realization
constraints are derived for the  standard  form and the
general form, respectively. A network theory is then developed for
synthesizing linear dynamical mixed quantum-classical stochastic
systems of the  standard  form in a systematic way. One feedback network
architecture is proposed for this network realization.

\section*{Acknowledgement}
The authors wish to thank the anonymous reviewers for their careful reading and constructive comments.

\end{document}